\documentclass[a4paper,11pt]{article}

\usepackage{hyperref}
\usepackage{amssymb}
\usepackage{amsfonts}
\usepackage{amsbsy}
\usepackage{amsmath}
\usepackage{longtable}
\usepackage{a4wide}
\usepackage{color}
\usepackage{rotating}

\hypersetup
{%
pdfauthor = {Johanna Knapp, Maximilian Kreuzer, Christoph Mayrhofer, Nils-Ole Walliser},
pdftitle = {Toric Construction of Global F-Theory GUTs},
pdfsubject = {String model building},
pdfkeywords = {F-theory; Compactifications;},
}

\begin{document}
\thispagestyle{empty}
\begin{flushright}
TUW-11-02\\
IPMU11-0006\\
HD-THEP-11-01
\end{flushright}
\vspace{1cm}
\begin{center}
{\LARGE Toric Construction of Global F-Theory GUTs}
\end{center}
\vspace{8mm}
\begin{center}
{\large Johanna Knapp${}^{\dagger\ast}$\footnote{e-mail: johanna.knapp AT ipmu.jp}, Maximilian Kreuzer${}^{\ast}$, Christoph Mayrhofer${}^{\ddagger}$\footnote{e-mail: c.mayrhofer AT thphys.uni-heidelberg.de },\\ and Nils-Ole Walliser${}^{\ast}$\footnote{e-mail: walliser AT hep.itp.tuwien.ac.at}}
\end{center}
\vspace{3mm}
\begin{center}
{\it $^{\ast}$Institut f\"ur Theoretische Physik, TU Vienna\\
Wiedner Hauptstrasse 8-10, 1040 Vienna, Austria
}
\end{center}
\begin{center}
{\it $^{\dagger}$ Institute for the Physics and the Mathematics of the Universe (IPMU)\\
The University of Tokyo, 5-1-5 Kashiwanoha, Kashiwa 277-8583, Japan
}
\end{center}
\begin{center}
{\it $^\ddagger$Institut f\"ur Theoretische Physik, Universit\"at Heidelberg,\\
Philosophenweg 19, D-69120 Heidelberg, Germany}
\end{center}
\begin{center}
\end{center}
\vspace{15mm}
\begin{abstract}
\noindent We systematically construct a large number of compact Calabi-Yau fourfolds which are suitable for F-theory model building. These elliptically fibered Calabi-Yaus are complete intersections of two hypersurfaces in a six dimensional ambient space. We first construct three-dimensional base manifolds that are hypersurfaces in a toric ambient space. We search for divisors which can support an F-theory GUT. The fourfolds are obtained as elliptic fibrations over these base manifolds. We find that elementary conditions which are motivated by F-theory GUTs lead to strong constraints on the geometry, which significantly reduce the number of suitable models. The complete database of models is available at \cite{data}. We work out several examples in more detail.

\end{abstract}
\vspace{2cm}
\begin{flushright} 
{\it In memory of  Maximilian Kreuzer}
\end{flushright} 
\vspace{1.2cm}
\newpage
\setcounter{tocdepth}{2}
\tableofcontents
\setcounter{footnote}{0}

\section{Introduction and Summary}
Starting with \cite{Donagi:2008ca,Beasley:2008dc,Beasley:2008kw}, F-theory has been recognized as a setup to elegantly construct Grand Unified Theories (GUTs) in string theory. The GUT model is localized on a seven-brane $S$ inside a complex three-dimensional manifold $B$ which is the base of a compact elliptically fibered Calabi-Yau fourfold $X_4$. Requiring a decoupling limit between gauge and gravity degrees of freedom makes it possible to discuss many questions in a gauge theory that captures the physics in the vicinity of the GUT brane $S$. These local F-theory GUTs have a rich yet simple structure which allows to analyze many phenomenological questions in remarkable detail. See for instance \cite{Heckman:2010bq} for a review. Due to the localization of gauge degrees of freedom on the seven-brane, in contrast to GUT theories coming from the heterotic string, F-theory provides a framework for a bottom-up approach to constructing realistic models from string theory. There, the first priority is to work out the phenomenological details of a model without worrying about the full string compactification. While the success of this approach speaks for itself, it is necessary to connect the bottom-up results with top-down constructions where the paradigm is to find a consistent string compactification which can ideally accommodate all the features of the local models. Finding and understanding global F-theory models has recently received increased attention. 

There are several reasons to consider a full F-theory compactification on an elliptically fibered Calabi-Yau fourfold. The obvious reason is of course that there are issues which cannot be addressed in local models, most notably monodromies, fluxes and anomaly cancellation. These questions have been addressed recently in \cite{Andreas:2009uf,Blumenhagen:2009yv,Marsano:2009wr,Grimm:2009yu,Blumenhagen:2010ja,Cvetic:2010rq,Hayashi:2010zp,Grimm:2010ez,Marsano:2010ix,Chung:2010bn,Cvetic:2010ky,Cecotti:2010bp,Marsano:2010sq,Collinucci:2010gz,Chiou:2011js}. Another motivation, which will be the central concern of this paper, is to explicitly construct compact Calabi-Yau fourfolds and to check whether they are suitable for F-theory model building. This is necessary in order to show whether the realistic models coming from a local construction have an embedding in a string compactification. Furthermore, we wanted to build a database of examples which contains the data necessary for GUT model building.

The main goal of this paper is to give a systematic construction of a particular class of fourfold geometries and to analyze them in view of F-theory model building. Since a full classification of Calabi-Yau fourfolds, including the subset of elliptically fibered ones, is not available we aim to provide a set of examples within a well-defined framework. Toric geometry is a valuable and versatile mathematical tool for constructing Calabi-Yau manifolds. A prescription to use toric geometry to construct global F-theory GUTs has been given in \cite{Blumenhagen:2009yv} and further elaborated on in \cite{Grimm:2009yu}. See also \cite{Weigand:2010wm} for a recent review article and \cite{Cvetic:2010rq} for a closely related construction. The general idea is the following: first, find a base manifold $B$ which is a blowup of a Fano hypersurface in $\mathbb{P}^4$. In a second step, obtain a Calabi-Yau fourfold by constructing an elliptic fibration over the base $B$. This Calabi-Yau is then a complete intersection of two hypersurfaces in a six-dimensional toric ambient space. In \cite{Chen:2010ts} a class of models has been worked out where the base manifold $B$ is a Fano hypersurface in $\mathbb{P}^4$ with up to three point or curve blowups. This extended the set of examples given in \cite{Blumenhagen:2009yv,Grimm:2009yu} but the geometries were still in a very restricted class. For instance, no examples in a general weighted projective space had been considered. In this article we will systematically construct this more general type of models. The present extension allows us, for example, to set up global F-theory GUTs on $dP_8$s that have not been found in the previous investigations.  

In order to find more general fourfold geometries we look at the construction of \cite{Blumenhagen:2009yv} from a slightly different point of view. Instead of considering blowups of Fano threefolds, we pick a subset of $1088$ of the $473\,800\,776$ reflexive polyhedra in four dimensions~\cite{Kreuzer:2000xy}. These polyhedra describe toric ambient spaces for Calabi-Yau threefolds. In contrast to looking at the Calabi-Yau case, we consider hypersurfaces in these toric ambient spaces that have homogeneous equations with multidegree smaller than in the Calabi-Yau case. This will define the base manifold $B$. The elliptically fibered Calabi-Yau fourfolds can be constructed from the base data using standard tools in toric geometry. In our computer-based search for models we have made extensive use of the software package PALP \cite{Kreuzer:2002uu}. In total we have found $569\,674$ base geometries.

Having constructed the geometries is only the first step of the program. Step two is to filter out those models which are usable in F-theory model building. Our goal was to formulate some elementary and general constraints that can be phrased in the toric language. These constraints can be divided up into conditions on the base geometry and conditions on the fourfold. While the former are specific to F-theory model building, the latter are of a more technical nature. As for the base manifolds, the first constraint is regularity. Hypersurfaces that are not Calabi-Yau may inherit the singularities of the toric ambient space. One sufficient criterion for regular hypersurfaces, which can be examined using toric methods, is base point freedom: given an empty base locus, any point-like singularity of the ambient space can be avoided by a generic choice of the hypersurface equation. We can impose further constraints on the toric divisors of the base $B$. Since we would like to construct F-theory models on these divisors, del Pezzo surfaces are particularly interesting. In local F-theory GUTs the del Pezzo condition guarantees a decoupling limit. Furthermore, certain vanishing theorems avoid exotic matter in $SU(5)$ GUTs \cite{Beasley:2008kw}. For global models decoupling limits are more subtle and yield further constraints on the base geometries. The conditions on the complete intersection Calabi-Yau fourfold are more elementary. In order to be able to use the tools of toric geometry, we restrict to those examples where the Calabi-Yau data is encoded in a reflexive lattice polytope and where the information about hypersurface equations is given by a nef partition. In our construction it is not automatic that the nef partition is compatible with the elliptic fibration over the base $B$. Another issue is the reflexivity of the polytope that encodes the toric data the fourfold. A majority of the fourfolds we have constructed is not described in terms of reflexive polytopes. Reflexivity is important for mirror symmetry but since this is not required in our setup Calabi-Yau fourfolds coming from non-reflexive polytopes may be interesting to look at. However, we lack several mathematical and computational tools to deal with them, which is why we have to exclude them in our discussion. Finally, there is unfortunately also a computational constraint: since the lattice polytopes for Calabi-Yau fourfolds can be quite large, a fair amount of models cannot be analyzed due to numerical overflows and long calculation times.\\
Having reduced the number of interesting models by the constraints above we can explicitly construct F-theory GUTs using the prescription of \cite{Blumenhagen:2009yv}. We will focus on $SU(5)$ and $SO(10)$ GUTs and analyze some basic properties such as genera of matter curves and the number of Yukawa couplings. We will also construct $U(1)$-restricted models as introduced in~\cite{Grimm:2010ez}.\\\\
This article is structured as follows. In section~\ref{sec-global} we review the toric construction and give a detailed explanation of the tools of toric geometry that are necessary to carry out the calculation. In section \ref{sec-analysis} we analyze the geometries we have constructed. Furthermore we discuss some examples and comment on the discrepancy of Euler numbers between the toric calculation and a formula given in \cite{Blumenhagen:2009yv}.  A match between the Euler numbers obtained from toric geometry and those obtained from the formula of \cite{Blumenhagen:2009yv} indicates that a local description of the gauge fluxes in terms of the spectral cover construction is plausible. 
Section \ref{sec-conc} is reserved for conclusions and outlook.\\\\
{\bf Acknowledgments:} We dedicate this work to our co-author and advisor Maximilian Kreuzer who passed away before this paper was completed. As his students we have greatly profited from his vast knowledge and dedicated support. This work would not have been possible without his contribution.\\
We would like to thank Timo Weigand for valuable comments on the manuscript.
Furthurmore, we thank Andreas Braun and Harald Skarke for many helpful discussions and comments. JK would like to thank LMU Munich and TU Vienna for hospitality, and Ralph Blumenhagen, Thomas Grimm, Benjamin Jurke and Emanuel Scheidegger for useful comments.\\
The work of JK was supported by World Premier International Research Center Initiative (WPI Initiative), MEXT, Japan. CM was supported in part by the DFG through TRR33 ``The Dark Universe'' and the Austrian Research Funds FWF under grant number I192-N16. The work of N-OW was supported by the FWF under grant P21239-N16 and I192-N16.


\section{Construction of Global Models}
\label{sec-global}
In this section we explain how to construct global F-theory models. In section \ref{sec-setup} we recall the basic structure of global F-theory GUTs. Section \ref{sec-toric} is devoted to a short self-contained review of aspects of toric geometry, focusing on the tools and objects we need for our calculations. In section \ref{sec-base} we describe how to systematically construct the base manifolds $B$ as hypersurfaces in toric ambient spaces. Furthermore we discuss the properties of GUT divisors in $B$. Finally, section \ref{sec-fourfold} is devoted to the elliptically fibered Calabi-Yau fourfolds. 
\subsection{Setup}
\label{sec-setup}
The class of global F-theory models, we aim to construct, have been first introduced in \cite{Blumenhagen:2009yv}. The Calabi-Yau fourfolds are complete intersections of two hypersurfaces in a six-dimensional toric ambient space. Schematically, these equations have the following form:
\begin{equation}
\label{cicydef}
P_{B}(y_i,w)=0\,,\qquad\qquad P_W(x,y,z,y_i,w)=0\,.
\end{equation}
The first equation only depends on the coordinates $(y_i,w)$ of the base of the fibration. Here we have singled out one coordinate $w$, indicating that the divisor $S$, defined by $w=0$, is wrapped by the seven-brane which supports the GUT theory. The second equation in \eqref{cicydef} defines a Weierstrass model, where $(x,y,z)$ are the coordinates of the $\mathbb P_{231}$ fiber. For this type of elliptic fibrations $P_W$ has a Tate form which is globally defined:
\begin{equation}
\label{tate}
P_W=x^3-y^2+xyz a_1+x^2z^2a_2+yz^3a_3+xz^4a_4+z^6a_6\,,
\end{equation}
where the $a_n(y_i,w)$ are sections of $K_B^{-n}$ and $x$ and $y$ are section of $K_B^{-2}$ and $K_B^{-3}$, respectively. Constructing a Tate model is only the first step on the way to a F-theory GUT model. In order for the divisor $w=0$ to support the desired gauge group the sections $a_n(y_i,w)$ have to have a particular structure. Via Kodaira's classification \cite{Kodaira} and Tate's algorithm \cite{tate} the base-coordinate dependent coefficients $a_i$ in the Tate equation must factorize in a particular way with respect to $w$. In the following we will focus on the gauge groups $SU(5)$ and $SO(10)$. For $SU(5)$ we must have:
\begin{equation}
\label{su5tate}
a_1=b_5w^0\quad a_2=b_4w^1\quad a_3=b_3w^2\quad a_4=b_2w^3\quad a_6=b_0w^5\,,
\end{equation}
An $SO(10)$ model is specified as follows:
\begin{equation}
\label{so10tate}
a_1=b_5w^1\quad a_2=b_4w^1\quad a_3=b_3w^2\quad a_4=b_2w^3\quad a_6=b_0w^5\,.
\end{equation}
The $b_i$s are sections of some appropriate line bundle over $B$ that have at least one term independent of $w$.\\
Matter arises along curves inside the base manifold at loci where a rank $1$ enhancement of the GUT group takes place. In $SU(5)$ F-theory GUTs the matter curves are at the following loci inside $S$:
\begin{eqnarray}
b_3^2b4-b_2b_3b_5+b_0b_5^3=0&\quad& \textrm{{\bf 5} matter}\quad \textrm{$SU(6)$ enhancement}\,,\nonumber\\
 b_5=0&\quad& \textrm{{\bf 10} matter}\quad \textrm{$SO(10)$ enhancement}\,.
\end{eqnarray}
The matter curves for the $SO(10)$ models are at:
\begin{eqnarray}
b_3=0&\quad& \textrm{{\bf 10} matter}\quad \textrm{$SO(12)$ enhancement}\,,\nonumber\\
b_4=0&\quad& \textrm{{\bf 16} matter}\quad \textrm{$E_6$ enhancement}\,.
\end{eqnarray}
Yukawa couplings arise at points inside $B$ where the GUT singularity has a rank $2$ enhancement. In $SU(5)$ models the Yukawa points sit at:
\begin{eqnarray}
 b_4=0 \cap b_5=0&\quad& {\bf 10\:10\:5} \textrm{ Yukawas}\quad \textrm{$E_6$ enhancement}\,, \nonumber\\ 
b_2^2-4 b_0 b_4=0 \cap b_3=0&\quad& {\bf 10\:\bar{5}\:\bar{5}} \textrm{ Yukawas} \quad \textrm{$SO(12)$ enhancement}\,.
\end{eqnarray}
In the $SO(10)$-case we have the following Yukawa couplings:
\begin{eqnarray}
b_3=0 \cap b_4=0&\quad& {\bf 16\:16\:10} \textrm{ Yukawas}\quad \textrm{$E_7$ enhancement} \,,\nonumber\\ 
b_2^2-4 b_0 b_4=0 \cap b_3=0&\quad& {\bf 16\:10\:10} \textrm{ Yukawas} \quad \textrm{$SO(14)$ enhancement}\,.
\end{eqnarray}
By constructing the base manifold $B$ and the elliptically fibered Calabi-Yau fourfold we are able to give explicit expressions for the quantities defined above. Furthermore, knowing the homology classes of divisors we can obtain intersection numbers and other topological data of the GUT brane, the matter curves and the Yukawa couplings. In order to make these calculations we make use of toric geometry. In the following subsections we will explain the necessary ingredients for these computations.


\subsection{Toric Geometry}
\label{sec-toric}
In this section we give a brief overview of the construction of toric varieties and their subvarieties in terms of lattice polytopes. We furthermore set the notation which we will need in the rest of the paper. The reader has a vast choice of existing literature on the subject, for example \cite{Danilov,Fulton,Cox} and the very comprehensive \cite{CoxKatz}, to name a few. In particular \cite{Kreuzer:2006ax} addresses the construction of Calabi-Yau hypersurfaces and complete intersections with a focus on issues related to string duality. For a more pedagogical approach see for instance \cite{Bouchard:2007ik}.   

Toric varieties can be thought of as generalizations of weighted projective spaces. We can construct a toric variety $X$ in terms of $r$ homogeneous coordinates, an exceptional set $Z_\Sigma$, and the group identification $(\mathbb{C^*})^{r-n}\times G$:
\begin{equation}
X=\left(\mathbb{C}^r-Z_\Sigma\right)/\left((\mathbb{C^*})^{r-n}\times G\right)\;.
\end{equation}
These building blocks are encoded in a fan $\Sigma$ that completely determines $X$. The fan is a finite collection of strongly convex (i.e.~they always have an apex) integral (i.e.~they are spanned by lattice vectors) polyhedral cones with their apex in the origin such that the following conditions are satisfied: 1) any face of a cone $\sigma\in\Sigma$ belongs to $\Sigma$; and 2) given two cones $\sigma,\,\tau\in\Sigma$, their intersection is again contained in $\Sigma$. Note that in general $\sigma$ and $\tau$ may have different dimensions. The $n$-{\textit skeleton} $\Sigma(n)\subset\Sigma$ denotes the set of $n$-dimensional cones. Consider the rays $\rho_j\in\Sigma(1)$. Each of them is generated by an integral vector $v_j$ in a $n$-dimensional lattice, which we call the N-lattice. The primitive vector $v_j$ spans from the origin towards the nearest point of the lattice along the direction of $\rho_j$. To each primitive vector $v_j$ we associate a homogeneous coordinate $z_j$ and a divisor $D_j=\{\left[z\right]\in X: z_j=0\}$. The group $(\mathbb{C^*})^{r-n}$ is hence determined by the $r-n$ weighted scalings ($i=1, \dots ,r-n$) 
\begin{equation}
\left(z_1, \dots , z_r \right)\longrightarrow \left(\lambda^{w_{i1}}z_1, \dots ,\lambda^{w_{ir}}z_r\right) \hspace*{5mm} \text{with} \hspace*{5mm} \sum_{j\leq r} w_{ij}v_j = 0\in N \hspace*{3mm}\text{and}\hspace*{3mm}\lambda \in \mathbb{C^*}\,,
\end{equation}
where $w_{ij}$ are the entries of a $r\times(r-n)$ matrix we refer to as \textit{weight matrix}. The finite abelian group $G\cong N/\text{span}(v_1, \dots, v_r)$ accounts for phase symmetries. It arises if the one-skeleton does not span the entire N-lattice. For example, let us consider a lattice $\hat{N}$ that is completely spanned by $\Sigma\left(1\right)$. Further, consider a refinement $N\supset\hat{N}$ such that $N\neq\text{span}\left(v_1, \dots ,v_r\right)$. Then we have $G\cong N/\hat{N}$. Furthermore the fan determines the exceptional set $Z_\Sigma$. This is the set of invariant points under the continuous group identification.
A subset of coordinates is allowed to vanish simultaneously, i.e. $z_{j_1}= \dots = z_{j_k}=0$ (or equivalently $D_{j_1}\cdot\,\dots\,\cdot D_{j_k}\neq 0$),  iff there exists a cone that contains the corresponding rays $\rho_{j_1}, \dots ,\rho_{j_k}\subset\sigma$. The exceptional set is the union of sets $Z_I$ with minimal index sets $I$ of rays for which there is no cone that contains them: $Z_\Sigma=\cup_I Z_I$.

A divisor $D$ is a codimension one subvariety of the toric ambient space and is defined by the formal sum $D=\sum_j a_j D_j$, where the $\{D_j\}$ are a finite set of irreducible divisors. Relevant properties of divisors can be rephrased in terms of the combinatorics between lattice points and cones \cite{Kreuzer:2006ax,Jaron}. In order to show these relations, we first need to define the dual lattice to N as $M=\text{Hom}(N,\mathbb{Z})$ with the canonical pairing $\langle ,\rangle$. A divisor $D$ is Cartier if to each maximal-dimensional cone $\sigma\in\Sigma(n)$ there exists a point $m_\sigma\in M$ such that the coefficient of the formal sum is $a_j=-\langle m_\sigma,v_j\rangle$ for all rays $\rho_j\in\sigma$. Furthermore to each Cartier divisor $D$ we associate a lattice polytope as follows
\begin{equation}
\Delta_D=\{m\in M_\mathbb{R} : \langle m, v_j\rangle\geq  -a_j \;\;\;\forall \;\rho_j\leq r \}\subset M_\mathbb{Q},
\end{equation}
where $M_\mathbb{R}$ and $M_{\mathbb{Q}}$ are the real and rational extensions of M, respectively. The corresponding line bundle $\mathcal{O}(D)$ is determined by the sections 
\begin{equation}
s_{\Delta_D}=\sum_{m\in \Delta_D}c_m \prod_j z_j^{\langle m,v_j\rangle}.
\end{equation}
The globally defined hypersurface polynomial is then:
\begin{equation}
f_{\Delta_D}=\sum_{m\in \Delta_D}c_m \prod_j z_j^{\langle m,v_j\rangle+a_j}\;.
\end{equation}
A Cartier divisor $D$ is base point free iff $m_\sigma\in\Delta_D$ for all $\sigma\in\Sigma(n)$. Further, a Cartier divisor $D$ is ample iff there is a bijection between vertices of $\Delta_D$ and $m_\sigma\in\Sigma(n)$.  
Consider $\Delta\subset M$ defining an ample Cartier divisor. In this case it can be shown that there is a uniquely associated fan to such a polytope: the normal fan $\Sigma_\Delta$. This is the fan of cones over the faces of the dual polytope $\Delta^\circ\in N_\mathbb{R}$ defined by    
\begin{equation}
\Delta^\circ=\{x\in N_\mathbb{R} : \langle m, x \rangle \geq -1 \;\;\;\forall\; m\in\Delta\}\;.
\end{equation}
A lattice polytope whose dual is again a lattice polytope is called reflexive. In our work we have considered toric ambient spaces from normal fans of reflexive polytopes. There are three reasons for this choice. First, these toric varieties have well understood singularity properties. Second, we know how to calculate their Hodge numbers in terms of combinatorial formulas due to the works \cite{batyrev1,Batyrev:1995ca}. Third, we have a classification scheme for reflexive polytopes up to dimension four \cite{Kreuzer:2000xy}.   

A toric variety $X_\Sigma$ is smooth iff all cones of $\Sigma$ are simplicial and basic (i.e. generated by a subset of the lattice basis). The normal fan of a given reflexive polytope will not generally satisfy these conditions. However, in our setup, we can always resolve singularities in toric spaces by subdivisions of their fan \cite{Billera,GKZ,Klemm:2004km}.
Take the polytope $\Delta^\circ\subset N$ with all its lattice points, and consider a star triangulation thereof, i.e.~a triangulation where the maximal simplices always contain the origin. The fan over the facets of this polytope depends on the particular star triangulation we have chosen.  
Then reflexivity implies that there are no singularities at codimension lower than four.
For a four-dimensional polytope, hence, there can be only point-like singularities. A hypersurface without fixed points can always be deformed to avoid this kind of singularities. Hence, for our setups, a base point free (Cartier) divisor is  smooth.

The intersection ring of a non-singular compact toric variety is given by the quotient ring
\begin{equation}
\mathbb{Z}\left[D_1, \dots ,D_r\right]/\langle I_{SR}\,,\;I_{lin}\,\rangle\;.
\end{equation}
Here $I_{SR}$ is the Stanley-Reisner ideal with relations of the type $D_{j_1}\cdot\,\dots\,\cdot D_{j_l}=0$ for elements of the minimal index set $I$. Furthermore one must mod out the ideal $I_{lin}$ generated by the linear relations $\sum_{j}\langle m, v_j\rangle D_j=0$.
The intersection ring of an embedded hypersurface is given by restricting the intersection ring of the ambient space to the divisor $D$ describing the hypersurface as follows:\footnote{By abuse of notation $D$ denotes the divisor as well as the associated Poincar\'e dual element of the cohomology.}
\begin{equation}
D_{j_1}\cdot \ldots \cdot D_{j_{n-1}}\vert_D=\int_D D_j\wedge\ldots\wedge D_{j_{n-1}} = \int_X D_{j_1}\wedge\ldots\wedge D_{j_{n-1}}\wedge D\;.
\end{equation}

We need the K\"ahler cone of the toric variety to determine the volumes of the divisors. With this information we will be able to make statements about the existence of a decoupling limit. 
We obtain it by starting from its dual, the Mori cone. The Mori cone is the cone of (numerically) effective curves. We determine it using the Oda-Park algorithm \cite{OdaPark,CoxKatz}, that has been implemented in an still unreleased version of the PALP code \cite{maxnils}. The extended PALP uses the SINGULAR \cite{DGPS} program to determine the intersection ring. The triple intersection numbers are then redirected to PALP to calculate the Mori cone. In what follows we approximate the K\"ahler cone of the embedded hypersurface by that of the ambient space. Since there could be more effective curves on the hypersurface than the induced ones, the K\"ahler cone of the hypersurface may be smaller than the one of the ambient space.

\subsubsection{Induced divisors}\label{sec-induced-divisor}
In our setup the base manifold is a divisor embedded in a toric ambient space. The reader may ask under which conditions and to which extent the homology of the hypersurface is induced from the homology classes of the toric ambient space. Indeed, not all toric divisors of the ambient space may induce a divisor on the hypersurface. For a Calabi-Yau hypersurface given by a reflexive polytope $\Delta^\circ$, this is the case if we have a divisor ${D_\textmd{int.}}_i$ obtained from points that lie in the interior of a facet of the polytope. To observe this, we consider the intersection product, on the CY hypersurface, of some ${D_\textmd{int.}}_i$ with divisors not coming from interior points,
\begin{equation}\label{eq:intersection-with-cy-equation}
 D_\textmd{CY}\cdot {D_\textmd{int.}}_i\cdot D_{j_1}\cdot \ldots\cdot D_{j_{n-2}}=n_{i\,j_1\ldots j_{n-2}}\,.
\end{equation}
We add to this equation intersection products of the form:
\begin{equation}\label{eq:inter-of-noninter}
 D_j\cdot {D_\textmd{int.}}_i\cdot D_{j_1}\cdot \ldots\cdot D_{j_{n-2}}=0\,,
\end{equation}
where the $D_j$ is a divisor that does not lie on the facet of the ${D_\textmd{int.}}_i$. This intersection is zero because the fan of the toric space is obtained from a maximal triangulation of the defining lattice polytope. Hence, divisors that lie in the interior of a facet intersect only divisors that also lie on that facet.
The lattice polytopes that we consider are reflexive. Thus, for each facet f$_i$ of the polytope we  have a point $m_{\textmd{f}_i}\in M$ in the dual lattice polytope with $\langle m_{\textmd{f}_i},p_j\rangle=-1$ for all points $p_j\in \textmd{f}_i$. From $m_{\textmd{f}_i}$ we obtain the principal divisor
\begin{equation}
 D_{m_{\textmd{f}_i}}=\sum_{p_j\in \textmd{f}_i} - D_j+\sum_{p_k \in \Delta^\circ\backslash \textmd{f}_i} \langle m_{\textmd{f}_i},p_k\rangle D_k\,.
\end{equation}
Since $D_\textmd{CY}=\sum_{p_k \in \Delta^\circ}  D_k$, we can add up \eqref{eq:intersection-with-cy-equation} and \eqref{eq:inter-of-noninter} to 
\begin{equation}
 D_{m_{\textmd{f}_i}}\cdot {D_\textmd{int.}}_i\cdot D_{j_1}\cdot \ldots\cdot D_{j_{n-2}}=-n_{i\,j_1\ldots j_{n-2}}\,.
\end{equation}
A principal divisor always has intersection number zero with any other divisor, hence, we obtain $n_{i\,j_1\ldots j_{n-2}}=0$. Therefore, the divisor ${D_\textmd{int.}}_i$ does not intersect with the Calabi-Yau hypersurface.

In the case of a hypersurface with a generic (multi) degree we cannot use the above M-lattice vector to prove that divisors obtained from interior points do not lie on the hypersurface. However, we may find another vector $m$ such that its principal divisor is the sum of the divisor of the hypersurface and the sum of toric divisors that do not come from points of the considered facet.

For the general hypersurface case not only divisors coming from interior points of facets may not induce a divisor but also others. For example, the  lower bound on the hypersurface degrees that we will consider below is that they include all homogeneous coordinates. At the bound we may encounter situations where one of the toric divisors has the same weight as the hypersurface. In this case all toric divisors that do not intersect the divisor showing up linearly in the hypersurface equation will not lie on the hypersurface.


\subsection{Base Manifolds}
\label{sec-base}
\subsubsection{Toric data for base manifolds}
In this section we introduce the class of base manifolds $B$ we will be working with. We will consider base geometries that are non-negatively curved hypersurfaces in a toric ambient space. We restrict to hypersurfaces with hyperplane class positive and strictly smaller than the class of the anti-canonical bundle of the ambient space. An interesting class of manifolds to look at would be Fano threefolds. However, as has been argued in \cite{Cordova:2009fg}, Fanos do not allow for a decoupling limit. We are thus forced to look for more general hypersurfaces. In \cite{Blumenhagen:2009yv,Grimm:2009yu,Chen:2010ts} such examples have been obtained by constructing point and curve blowups of those Fano threefolds which are hypersurfaces in $\mathbb{P}^4$. A systematic construction for up to three point and curve blowups has been undertaken in \cite{Chen:2010ts} by a classification of the weight systems specifying the toric ambient space. What we would like to achieve here is to construct base manifolds in a more general class of ambient spaces, using toric geometry. In order to do so we will use a slightly different point of view than in \cite{Chen:2010ts}: instead of classifying weight systems corresponding to blowups we will specify the ambient space by reflexive polyhedra in four dimensions. These have been classified in \cite{Kreuzer:2000xy}. Since we are not looking for Calabi-Yaus each of these polytopes will give us a large number of models since there are typically many possibilities to define hypersurfaces   inside the ambient space defined by the polytope that fulfill the above above hyperplane class constraint. Therefore it has not been possible for us to construct base manifolds from all the $473\,800\,776$ reflexive polyhedra in four dimensions. Instead, we will look at a class of geometries specified by N-lattice polytopes which define toric ambient spaces which are fourfolds with Picard number less than five. Concretely, we have looked at N-lattice polytopes with up to nine points, including the origin. Not all the points of a polytope are also vertices. We have divided up the data accordingly. This is summarized in Table \ref{tab-poly}. The polytope data can be recovered from this information at \cite{cydata}.
\begin{table}
\begin{center}
\begin{tabular}{c|c|c}
\# of points & \# of vertices & \# of polytopes\\
\hline
$6$&$5$&$3$\\
$7$&$5$&$7$\\
$7$&$6$&$18$\\
$8$&$5$&$9$\\
$8$&$6$&$70$\\
$8$&$7$&$89$\\
$9$&$5$&$13$\\
$9$&$6$&$115$\\
$9$&$7$&$406$\\
$9$&$8$&$358$\\
\hline\hline
&&$1088$\\
\end{tabular}\caption{Lattice polytopes specifying toric ambient spaces for $B$}\label{tab-poly}
\end{center}
\end{table}
The points of the N-lattice polytopes encode the weight matrices which we can recover using PALP. The next step in constructing the base manifolds is to specify a hypersurface of degrees $d_i$, where $i$ runs over the rows in the weight matrix. The type of hypersurface we are interested in constrains the number of possible degrees. If  $d_i=\sum_jw_{i,j}$, where $w_{ij}$ are the homogeneous weights of the variables, the hypersurface will be Calabi-Yau. This gives an upper bound for the degrees: for our purposes we have to consider hypersurface degrees such that at least one of the $d_i$ is strictly smaller than the sum of the weights. Furthermore, we would like our base manifold $B$ to be a genuine codimension $1$ hypersurface inside the toric ambient space. Therefore we impose the condition that each variable has to appear in at least one monomial of the hypersurface equation. If the homogeneous weight of a variable is higher than the hypersurface degree the variable will certainly not appear in the hypersurface equation. This gives a lower bound on the hypersurface degree. Since this bound is necessary but not sufficient, one has to check for each model if indeed all the variables appear in the hypersurface equation. For the ambient spaces specified by the $1088$ polytopes above we have constructed all the hypersurfaces satisfying these conditions. In this way we have obtained as many as $569\,674$ potential candidates for bases of an F-theory compactification.

\subsubsection{GUT data from base manifolds}
Even though we are ultimately interested in constructing a full F-theory compactification on a Calabi-Yau fourfold, a lot of important information about the GUT model is already encoded in the geometry of the base manifold. What is more, in many cases this data can be inferred from the toric data of the ambient space. In the following we discuss what we can learn from the geometry of $B$ and how to compute phenomenologically relevant data using toric geometry. In our discussion about the GUT brane $S$, which wraps a toric divisor in $B$, we will focus on $SU(5)$ and $SO(10)$ models. 
\subsubsection*{Singularities}

Singularities can either come from singularities of the ambient space or the hypersurface equation. Since the ambient space of the base manifold is characterized by a reflexive polytope in four dimensions, only point-like singularities arise there. On the other hand the hypersurface itself can be singular. A hypersurface given by an equation $W(x_1,\ldots x_n)=0$ is singular at a locus $x_{sing}$ if:
\begin{equation}
\label{sing-cond}
W|_{x_{sing}}=0\quad \partial_{x_i}W|_{x_{sing}}=0\qquad x_{sing}\in X_6
\quad i=1,\ldots,N\,.
\end{equation}

A sufficient condition for regularity is that the divisor defining the hypersurface is base point free. In this case the hypersurface can be transversally deformed in every point. By Bertini's theorem, it will not have any singularities of the kind of~\eqref{sing-cond}. Additionally, the hypersurface will miss possible point singularities of the ambient space which are the only singularities of our toric ambient spaces of $B$. 
The base point free condition is given purely in terms of the combinatorics of the lattice polytope and therefore quite simple to check.

\subsubsection*{Almost Fano manifolds}
An almost Fano threefold is an algebraic threefold that has a non-trivial anti-canonical bundle with at least one non-zero section at every point.
Our toric construction of base manifolds does not necessarily lead to almost Fano manifolds. Thus, we check this criterion by explicitly searching for non-zero sections in every example. In the examples analyzed in \cite{Chen:2010ts} a connection between the almost Fano property of $B$ and the reflexivity of the lattice polytope associated to the elliptically fibered fourfold had been observed. 
\subsubsection*{Del Pezzo Divisors}
Having specified a base manifold $B$, the next task is to identify suitable GUT divisors $S$. For this purpose we will systematically search for del Pezzo divisors inside $B$. There are several motivations to look for del Pezzos. In local F-theory GUTs the del Pezzo property ensures the existence of a decoupling limit \cite{Beasley:2008dc,Beasley:2008kw}. For $SU(5)$ GUT models, the fact that del Pezzos have $h^{0,1}=h^{2,0}=0$ implies some powerful vanishing theorems which forbid exotic matter after breaking $SU(5)$ to the Standard Model gauge group \cite{Beasley:2008kw}. However, one should keep in mind that there are other possibilities besides del Pezzos: as pointed out in \cite{Donagi:2008ca}, for the F-theory model to have a heterotic dual $S$ may also be a Hirzebruch or an Enriques surface. Recently, a construction of an F-theory GUT on an Enriques surface has been discussed \cite{Braun:2010hr}.\\
We will identify candidates for del Pezzo divisors inside $B$ by their topological data. All the calculations can be done using toric geometry. Suppose the base manifold has hyperplane class which, by abuse of notation, we also call $B$ and is embedded in a toric ambient space with toric divisors $D_i$. The total Chern class of a particular divisor $S$ in $B$ is:
\begin{equation}
c(S)=\frac{\prod_i(1+D_i)}{(1+B)(1+S)}
\end{equation}
A necessary condition for the divisor $S$ to be $dP_n$ is that it must have the following topological data:
\begin{equation}
\label{dpcond1}
\int_Sc_1(S)^2=9-n\qquad \int_Sc_2(S)=n+3\qquad\Rightarrow\qquad \chi_h=\int_S\mathrm{Td}(S)=1,
\end{equation}
where $\chi_h$ is the holomorphic Euler characteristic.
Since del Pezzos are Fano twofolds, we have a further necessary condition. The integrals of $c_1(S)$ over all torically induced curves\footnote{Of course positivity should hold for all curves, but within the framework of toric geometry we can only verify this for the divisors induced from the ambient space.} on $S$ have to be positive:
\begin{equation}
\label{poscurve}
D_i\cap S\cap c_1(S) > 0\qquad D_i\neq S \qquad \forall D_i\cap S\neq\emptyset\,.
\end{equation}
\subsubsection*{Genus of matter curves}
Assuming that we have set up the right GUT theory on the divisor S,
matter is localized at curves of further enhancement of the singularity.  The curve classes $M$ of the matter curves can be expressed in terms of the toric divisors of the ambient space. The genus of the matter curve can be computed using its first Chern class and the triple intersection numbers. The total Chern class is:
\begin{equation}
c(M)=\frac{\prod_i(1+D_i)}{(1+B)(1+S)(1+M)}
\end{equation}
After expanding this expression to obtain $c_1(M)$, the Euler number can be calculated by the following intersection product:
\begin{equation}\label{eq:euler-number-curves}
\chi(M)=2-2g(M)=c_1(M)\cap M\cap S
\end{equation}
Note that we have made the assumption that the matter curves are generic and do not factorize. This may not always be the case and then formula \eqref{eq:euler-number-curves} will yield the sum of the Euler numbers of the factorized curves as result. This may for instance lead to negative values for the genus of the matter curve if we na\"ively assume a single connected curve. 
The genus of $M$ gives us information about the number of moduli on the matter curve. Since these moduli will eventually have to be stabilized, matter curves of low genus are desirable from a phenomenological point of view.
\subsubsection*{Yukawa Points}
Yukawa couplings arise at points inside $B$ where the GUT singularity has a rank $2$ enhancement. In the generic situation the equations specifying the Yukawa points can be expressed as classes $Y_1,Y_2$ in terms of the toric divisors. The number of Yukawa points is then given by the following intersection product:
\begin{equation}
\label{nyuk}
n_{\textrm{Yukawa}}=S\cap Y_1\cap Y_2
\end{equation}
In order to account for the Standard Model Yukawa couplings only a small number of Yukawa points is needed. In $SO(10)$-models, for example, all the Standard Model couplings descend from ${\bf 16\:16\:10}$ Yukawas, which is why it would be nice to find a geometry where the number of ${\bf 16\:10\:10}$ Yukawa points is as small as possible. Most of the known global geometries come with a large number of Yukawa points. The situation is particularly bad for $dP_n$ with small $n$ \cite{Hayashi:2009bt}. Our analysis shows however that $dP_0$ and $dP_1$ are by far the most common del Pezzo divisors in the base manifolds.
\subsubsection*{Decoupling limit}
One of the key issues which allows for the discussion of GUT models within F-theory locally around the seven-branes is the existence of a decoupling limit. The Planck mass and the mass scale of the GUT theory are related to the geometry in the following way:
\begin{equation}
M_{pl}^2\sim \frac{M_s^8}{g_s^2}\mathrm{Vol}(B)\qquad M_{GUT}\sim \mathrm{Vol}(S)^{-\frac{1}{4}}\qquad 1/g^2_\textmd{YM}\sim \frac{M_s^4}{g_s}\mathrm{Vol}(S)\,,
\end{equation}
see for instance~\cite{Cvetic:2011vz}. Therefore one has:
\begin{equation}
\frac{M_{GUT}}{M_{pl}}\sim g^2_\textmd{YM}\frac{\mathrm{Vol}(S)^{3/4}}{\mathrm{Vol}(B)^{1/2}}
\end{equation}
There are two ways to achieve a small value for $M_{GUT}/M_{pl}$. These are often referred to as the physical and the mathematical decoupling limit. In the physical decoupling limit the volume of the $GUT$ brane $S$ is kept finite while $\mathrm{Vol}(B)\rightarrow\infty$. The mathematical decoupling limit takes $\mathrm{Vol}(S)\rightarrow 0$ for finite volume of $B$.  In the case of a rigid del Pezzo divisor the mathematical decoupling limit should always be possible. Thus, it can be used to check whether a del Pezzo is rigid.
Here we study the dependence of the volumes of $S$ and $B$ in terms of the K\"ahler moduli. This discussion tells us if a decoupling limit can in principle be realized in the given geometry. If the limits are actually realized is a question of moduli stabilization, which we will not discuss here. \\
The question of whether there exists a decoupling limit can again be addressed within the realm of toric geometry. In order to obtain positive volumes we must find a basis of the K\"ahler cone. The K\"ahler cone of the hypersurface describing the base is hard to compute. Therefore we will approximate it by the K\"ahler cone of the ambient space. Having found a basis $K_i$ of the K\"ahler cone, the K\"ahler form $J$ can be written as $J=\sum_ir_iK_i$ with $r_i>0$. Using the Mori cone we can express $K_i$ in terms of the toric divisors $D_i$. The triple intersection numbers restricted to $B$ allow us to compute the following volumes in terms of the K\"ahler parameters $r_i$:
\begin{equation}
\mathrm{Vol}(B)=J^3\qquad \mathrm{Vol}(S)=S\cdot J^2
\end{equation}
The existence of a mathematical and physical decoupling limits can be deduced from the moduli dependence of these volumes.  As was first observed in \cite{Grimm:2009yu} these two decoupling limits may be governed by different vectors in the K\"ahler cone.
\subsection{Elliptically Fibered Calabi--Yau Fourfolds}
\label{sec-fourfold}
\subsubsection{Construction of the Fourfolds}
\label{sec-fourfold-constr}
We now go on to construct an elliptically fibered Calabi-Yau fourfold from $B$. We obtain such an elliptic fibration by first fibering $\mathbb{P}_{231}[6]$ over the toric ambient space of the base manifold. Thus, we extend the weight matrices describing the ambient space of $B$ by suitable weights for the new fiber coordinates $(x,y,z)$. This is done such a way that $x$, $y$, and $z$ transform  as $K_B^{-2}$, $K_B^{-3}$, and $\mathcal O_B$, respectively. We also add an extra weight vector $(2,3,1,0,0,\ldots,0)$ to account for the $\mathbb{P}_{231}$.
In order to have a well defined torus fibration, the coefficients $a_i$ of equation \eqref{tate} have to be sections of $K_B^{-n}$ with some appropriate power $n$. The sums of the degrees of the hypersurface equation of the base and of the equation specifying the elliptic fibration are now equal to the degree of the anti-canonical bundle of the ambient toric sixfold. Hence, the complete intersection of these two equations is a Calabi-Yau manifold. This variety may be singular in some cases. 
The complete intersection Calabi-Yaus we consider here are given in terms of a pair of reflexive lattice polytopes $\Delta$ and $\Delta^{\circ}$, together with a nef partition:
\begin{eqnarray}                                        \label{cicy-nef}
        \Delta=\Delta_1+\ldots+\Delta_r&& \Delta^{\circ}
        =\langle \nabla_1,\ldots,\nabla_r\rangle_{\mathrm{conv}}\nonumber\\
                &(\nabla_n,\Delta_m)\geq-\delta_{nm}&\\
        \nabla^{\circ}=\langle \Delta_1,\ldots,\Delta_r\rangle_{\mathrm{conv}}
        &&\nabla=\nabla_1+\ldots+\nabla_r\nonumber
\end{eqnarray}
Here, $\langle\ldots\rangle_{\mathrm{conv}}$ denotes a convex hull of lattice polytopes, and $\Delta=\Delta_1+\ldots+\Delta_r$ (and analogously for $\nabla$) is a Minkowski sum.

The extension of the weight systems of the base threefold is straight forward. However, there are several issues of both conceptual and technical nature which prevent us from constructing an F-theory compactification for every base $B$. These are discussed in the following. 

\subsubsection*{Software Constraints}
There are two main constraints affecting our search for complete intersection Calabi-Taus (CICYs). First, PALP was originally designed to analyze complete intersection Calabi-Yaus of the type \eqref{cicy-nef}, which does not cover all the possibilities we encounter in our construction of global F-theory GUTs. The software efficiently analyzes combined weight systems to find their description in terms of (six-dimensional) reflexive polytopes. Afterwards PALP determines their nef partitions and the Hodge numbers of the CICY. Given a six-dimensional reflexive polytope describing the ambient space, the common zero locus of any two transversal equations is a suitable Calabi-Yau. Note that the two defining equations do not have to descend from the nef partitions, but only for nef partitions it is known how to determine the Hodge numbers of the CICY in terms of combinatorial data \cite{Batyrev:1995ca}. Thus, we could only do detailed calculations for examples that fulfill the requirements of \eqref{cicy-nef}. In fact, not all of the combined weight systems we have constructed extending the base weight matrices correspond to reflexive polytopes or do have nef partitions. Table \ref{tab-fourfold-goodbad} in section \ref{sec-analysis} shows how many CICYs satisfy these conditions. Reflexivity has turned out to be a severe constraint. 

The second obstacle in our analysis of the fourfolds is that due to computational constraints we have not been able to determine the six-dimensional N polytopes for all weight matrices. The last column of table \ref{tab-fourfold-goodbad} shows where the software has failed. The entries in the columns give information of two types of errors that can occur when determining the polytopes in the N-lattice: in most cases the error comes from the the issue that PALP cannot determine the N-lattice polytope by solving the equations encoded in the weight matrices. This problem might in principle be overcome by choosing the points of the N-lattice polytope as an input instead of the weight matrix. In fewer cases the N-lattice polytope can be found but an upper bound to the number of points is violated. The upper bound could be increased but that usually leads to very long computation times. The error distribution is in agreement with the intuitive idea that the complexity of the weight matrices increases with the number of points. For the fibrations over polytopes with $8$ points where $7$ of which are vertices we get an error in $10,9\%$ of the cases, for polytopes with $9$ points and $8$ vertices we have an error occurrence of $28,5\%$.       

The fourfold data available at \cite{data} do not contain the Hodge numbers of the CICYs. They can be easily determined with help of the nef-function of PALP.\footnote{In fact \texttt{nef.x} yields the Hodge numbers by default. The flag \texttt{-p} deactivates their calculation. For more details we refer to the help information: \texttt{nef.x -h}.} However, due to the complexity of the polytopes their calculation would have been too time consuming to be applied to every model we had.

\subsubsection*{Compatibility with the Elliptic Fibration}
Once we have found a Calabi-Yau fourfold characterized by a pair of dual polyhedra and its nef partitions, we still need to make sure that one of the nef-partitions is compatible with the desired elliptic fibration. The most elementary requirement for a well-defined Weierstrass model is of course that the points in $\Delta^{\circ}$ corresponding to the coordinates of the torus fiber are all in the same component of the nef-partition. However, this criterion is not sufficient in order to recover the desired Weierstrass model. We also have to make sure that the coefficients $a_n$ in~\eqref{tate} transform appropriately as sections of $K_B^{-n}$. This translates into conditions on the on the (sums of) weights of the variables in the individual nef partitions. 

\subsubsection{Engineering GUT models}
By now we have constructed complete intersection Calabi-Yau fourfolds of type (\ref{cicydef}). The next step is to obtain a GUT model. This is achieved by imposing the factorization constraints such as (\ref{su5tate}) or (\ref{so10tate}) on the coefficients $a_r(y_i,w)$ in the Tate equation (\ref{tate}). The procedure can be done within the toric framework, as has been proposed first in \cite{Blumenhagen:2009yv}. The hypersurface constraints can be recovered from the toric data as follows:
\begin{equation}
f_m=\sum_{w_k\in\Delta_m}c_k^{m}\prod_{n=1}^{2}\prod_{\nu_i\in\nabla_n}x_i^{\langle\nu_i,w_k\rangle+\delta_{mn}}\qquad m,n=1,2\,,
\end{equation}
where the $c_k^{m}$ are complex structure parameters. The Tate form~(\ref{tate}) implies that the $a_n$ appear in the monomials which contain $z^n$. We can isolate these monomials by identifying the vertex $\nu_z$ in $(\nabla_1,\nabla_2)$ that corresponds to the $z$-coordinate.  All the monomials that contain $z^r$ are then in the following set:
\begin{equation}
A_r=\{w_k\in\Delta_m\::\:\langle\nu_z,w_k\rangle-1=r\}\qquad \nu_z\in\nabla_m,
\end{equation}
where $\Delta_m$ is the dual of $\nabla_m$, which denotes the polytope containing the $z$-vertex. The polynomials $a_r$ are then given by the following expressions:
\begin{equation}
a_r=\sum_{w_k\in A_r}c_k^{m}\prod_{n=1}^2\prod_{\nu_i\in\nabla_n}y_i^{\langle \nu_i,w_k\rangle+\delta_{mn}}\vert_{x=y=z=1}
\end{equation}
Now we can remove all the monomials in $a_r$ which do not satisfy the factorization constraints of the singularity classification. In order to perform this calculation we have to identify the fiber coordinates $(x,y,z)$ and the GUT coordinate $w$ within the weight matrix of the fourfold.

The restriction to a specific GUT group amounts to removing a considerable amount of M-lattice points. As has been observed in \cite{Chen:2010ts} these manipulations may destroy the reflexivity of the polytope. The dual polytope in the N-lattice will have acquired additional points that can be interpreted as exceptional divisors obtained by blowing up the GUT singularity \cite{Blumenhagen:2009yv,Grimm:2009yu}. 

\subsubsection*{$U(1)$-restricted models}
Recently there has been active discussion in the literature on how to globally define fluxes in F-theory models. While a full answer to this problem is still unknown there has been some progress in incorporating the spectral cover construction into global models \cite{Grimm:2010ez,Marsano:2010ix}. For phenomenological reasons one has to make sure that, in $SU(5)$ models, the spectral cover splits. This is necessary to forbid dimension four proton decay operators. In $SO(10)$ models a split spectral cover is used to generate chiral fermions \cite{Chen:2010tp,Chen:2010ts}. However, as has been argued in \cite{Hayashi:2010zp,Grimm:2010ez} the local picture of a split spectral cover may in general not be sufficient. The authors of \cite{Grimm:2010ez} have shown that a lift of the local split spectral cover construction to a globally defined ``$U(1)$-restricted Tate models'' can give the needed further selection rule. This is achieved by imposing a global $U(1)_X$ symmetry in the elliptic fibration. In terms of the Tate model this is achieved by setting $a_6=0$. In the toric language this corresponds to removing even more points in the M-lattice, in addition to the manipulations needed for imposing the GUT model. Due to this procedure the Euler number decreases significantly, which is problematic for tadpole cancellation. Since the $U(1)$-restriction removes even more points from the M-lattice, reflexivity might not be maintained.


\section{Data Analysis}
\label{sec-analysis}
In this section we analyze our data.\footnote{The complete data concerning the base manifolds, their analysis, as well as the elliptically fibered fourfolds and the GUT models is available at \cite{data}. For details on the data format we refer to the \texttt{README.txt} file the reader can find there.} In total we have produced $569\,674$ base geometries. We will discuss their properties and the associated elliptically fibered fourfolds.

\subsection{Base Manifolds}
We collect the information about the base geometries in several tables. Our discussion will be concerned with properties of the base manifold, properties of its divisors and furthermore matter curves, Yukawa couplings as well as the existence of a decoupling limit.\\
In table \ref{tab-base-summary} we summarize some information about the base geometries. We subdivide the models into classes p$n$v$m$, denoting models based on polytopes which have $n$ points and $m$ vertices. The last three columns in the table indicate how many of the base manifolds are Cartier divisors, base point free or almost Fano.
\begin{table}
\begin{center}
\begin{tabular}{c|c|c|c|c|c}
class & \# of polytopes & \# of base manifolds & Cartier & BP-free & almost Fano\\
\hline\hline
p6v5&$3$&$12$&$6$&$6$&$10$\\
p7v5&$7$&$155$&$66$&$31$&$39$\\
p7v6&$18$&$307$&$199$&$131$&$94$\\
p8v5&$9$&$812$&$424$&$86$&$73$\\
p8v6&$70$&$6691$&$3265$&$816$&$584$\\
p8v7&$89$&$8168$&$4464$&$1542$&$779$\\
p9v5&$13$&$8238$&$1243$&$77$&$155$\\
p9v6&$115$&$84848$&$27037$&$1651$&$1542$\\
p9v7&$406$&$257024$&$107119$&$10515$&$5955$\\
p9v8&$358$&$203419$&$101562$&$14564$&$5677$\\
\hline
\hline
{\bf total}&$1088$&$569674$&$245385$&$29401$&$14908$\\
\hline
\end{tabular}\caption{Analysis of the base manifolds.}\label{tab-base-summary}
\end{center}
\end{table}
We note that base point freedom and in particular the almost Fano property are extremely rare items. As for almost Fano, it turns out that this property of the base manifold is not needed in order to have a Calabi-Yau fourfold that is characterized by a reflexive polytope.\\
In our search for geometries that are suitable for F-theory model building we have focused on identifying del Pezzo divisors inside the base manifold. The results of our search are summarized in table \ref{tab-dp-summary}. All the divisors in this counting satisfy (\ref{dpcond1}) and (\ref{poscurve}).  Among all the base geometries, we have identified $269\,636$ models with del Pezzo divisors, and a total number of $471\,844$ del Pezzos. The $dP_n$ with $n=0,1,2$  are the most common ones.
\begin{sidewaystable}
\centering
\begin{tabular}{c|c|c|c|c|c|c|c|c|c|c|c}
class & base manifolds with dPs & \# of $dP_n$ & $dP_0$ & $dP_1$ & $dP_2$ & $dP_3$ & $dP_4$ & $dP_5$ & $dP_6$ & $dP_7$ & $dP_8$\\ 
\hline\hline
p6v5&$6$&$25$&$9$&$6$&-&-&-&-&$6$&$4$&-\\
p7v5&$66$&$150$&$36$&$72$&$4$&-&$2$&$1$&$17$&$14$&$4$\\
p7v6&$206$&$597$&$121$&$239$&$35$&$11$&$17$&$9$&$73$&$64$&$28$\\
p8v5&$429$&$787$&$133$&$431$&$43$&-&$14$&$8$&$75$&$45$&$38$\\
p8v6&$3322$&$6259$&$1074$&$2883$&$539$&$157$&$164$&$171$&$520$&$458$&$293$\\
p8v7&$4888$&$11449$&$1868$&$4162$&$1325$&$670$&$451$&$532$&$931$&$947$&$563$\\
p9v5&$3213$&$5415$&$1562$&$1740$&$274$&$61$&$115$&$31$&$617$&$949$&$66$\\
p9v6&$31160$&$45039$&$8598$&$20261$&$4228$&$1167$&$992$&$1023$&$3763$&$3823$&$1184$\\
p9v7&$113364$&$181672$&$31926$&$72056$&$22238$&$9432$&$5812$&$6632$&$12061$&$13839$&$7376$\\
p9v8&$112982$&$220451$&$35669$&$73549$&$32191$&$18130$&$11098$&$11394$&$14950$&$15183$&$8887$\\
\hline\hline
{\bf total}&$269636$&$471844$&$80996$&$175399$&$60877$&$29628$&$18665$&$19801$&$33013$&$35326$&$18439$\\
\hline
\end{tabular}\caption{Results of the del Pezzo analysis}\label{tab-dp-summary}
\end{sidewaystable}
So far, our discussion has included all possible choices of base manifolds. We can now collect those models which have some attractive features. For that reason we will now focus on those models where $B$ is regular and has at least one del Pezzo divisor that allows for a mathematical or physical decoupling limit. This leaves us with only a small fraction of models, as indicated in table \ref{tab-goodbase}. In the first column we count the number of models where the hypersurface divisor of $B$ is Cartier and there is at least one del Pezzo divisor with a mathematical or physical decoupling limit. In the second column we furthermore implement the constraint that $B$ is base point free. In the third column we count the total number of all del Pezzos (also those without decoupling limit) in the base point free geometries, where at least one dP-divisor allows for a decoupling limit.
\begin{table}
\begin{center}
\begin{tabular}{c|c|c|c}
{\bf class}&{\bf Cartier+dec+dP}&{\bf BP-free+dec+dP}&{\bf \# dP for BP-free+dec}\\
\hline
p6v5&-&-&-\\
p7v5&$29$&$22$&$74$\\
p7v6&$85$&$72$&$277$\\
p8v5&$224$&$74$&$212$\\
p8v6&$1492$&$665$&$2073$\\
p8v7&$2412$&$1264$&$4490$\\
p9v5&$726$&$62$&$239$\\
p9v6&$10900$&$1332$&$3334$\\
p9v7&$46142$&$8933$&$26776$\\
p9v8&$53356$&$13108$&$50930$\\
\hline\hline
{\bf total}&$115366$&$25532$&$88405$\\
\hline
\end{tabular}\caption{Base manifolds with del Pezzo divisors and decoupling limit.}\label{tab-goodbase}
\end{center}
\end{table}
\subsection{Fourfolds}
In this section we discuss the Calabi-Yau fourfolds which are elliptic fibrations over the base threefolds. The toric data of the fourfolds is obtained by extending the weight matrices associated to the base manifolds, as discussed in section \ref{sec-fourfold-constr}. Complete intersection Calabi-Yaus can be analyzed by PALP. The fourfold data contains a lot of information which is relevant for finding global F-theory GUT models. We can use the data to answer the following questions:
\begin{enumerate}
\item Does the extension of the weight matrix of the base lead to a reflexive polytope?
\item How many of the Calabi-Yau fourfolds have nef partitions that are compatible with the elliptic fibration over $B$?
\item Do the ``good'' base manifolds (i.e. those which are regular, have del Pezzo divisors and a decoupling limit) always extend to Calabi-Yau fourfolds, which are described in terms of reflexive polytopes and nef partitions?
\item After imposing a GUT group using the construction of \cite{Blumenhagen:2009yv}, are the fourfold polytopes still reflexive?
\item Does imposing the GUT model lead to further non-abelian enhancements on divisors other than the GUT divisor?
\item Can we implement a $U(1)$-restricted Tate model in order to impose a global $U(1)$--symmetry \cite{Grimm:2010ez} without destroying desirable properties on the Calabi-Yau fourfold?
\end{enumerate}
Even though we have the tools to answer all these questions, working out the details for a large class of models is quite tricky and takes up a lot of computing time. This is why we will address some of these issues, in particular the fifth question, only in several examples. \\\\
We start by answering the first question above. As a somewhat surprising outcome, only a very small fraction of threefold base manifolds can be extended to a Calabi-Yau fourfold which is described by a pair of reflexive polyhedra and at least one nef partition. We have found $27\,345$ such models. The results are summarized in table \ref{tab-fourfold-goodbad}. About one quarter of the extended weight systems could not be analyzed due to their complexity. 
\begin{table}
\begin{center}
\begin{tabular}{c|c|c|c|c}
{\bf class (base)}&{\bf reflexive+nef part.}&{\bf reflexive, no nef part.}&{\bf non-reflexive}&{\bf PALP errors}\\
\hline
p6v5&$10$&-&$2$&-\\
p7v5&$65$&$6$&$84$&-\\
p7v6&$128$&$7$&$172$&-\\
p8v5&$197$&$103$&$308$&$188+16$\\
p8v6&$1170$&$344$&$4481$&$660+36$\\
p8v7&$1051$&$267$&$5958$&$892+0$\\
p9v5&$256$&$146$&$583$&$7187+66$\\
p9v6&$4033$&$3530$&$61211$&$14861+1213$\\
p9v7&$12101$&$8963$&$176598$&$58439+928$\\
p9v8&$8334$&$5266$&$131835$&$57918+66$\\
\hline\hline
{\bf total}&$27345$&$18632$&$381232$&$140145+2325$
\end{tabular}\caption{Fourfold Polytopes}\label{tab-fourfold-goodbad}
\end{center}
\end{table}
For the rest of the discussion we will focus on those fourfolds which can be characterized by reflexive polytopes and have at least one nef-partition. At first we merge the fourfold data with the data of the base manifold in order to check how many of the ``good'' base manifolds also lead to Calabi-Yau fourfolds that are characterized by reflexive polytopes with nef partitions. Our findings are collected in table \ref{tab-fourfold-goodbase}. The number of models which have a reflexive fourfold polytope, where the base is regular and there is at least one del Pezzo divisor with a mathematical and/or physical decoupling limit is $7386$.
\begin{table}
\begin{center}
\begin{tabular}{c|c|c|c}
{\bf class}&{\bf $CY_4$+refl+nef}&{\bf Cartier base+$dP_n$+dec.}&{\bf BP-free base+$dP_n$+dec.}\\
\hline
p6v5&$10$&-&-\\
p7v5&$65$&$24$&$18$\\
p7v6&$128$&$61$&$57$\\
p8v5&$197$&$94$&$38$\\
p8v6&$1170$&$685$&$402$\\
p8v7&$1051$&$760$&$591$\\
p9v5&$256$&$5$&-\\
p9v6&$4033$&$1679$&$414$\\
p9v7&$12101$&$6909$&$2714$\\
p9v8&$8334$&$5794$&$3152$\\
\hline\hline
{\bf total}&$27345$&$16011$&$7386$\\
\hline
\end{tabular}\caption{CY fourfolds where the base manifolds are suitable for F-theory model building.}\label{tab-fourfold-goodbase}
\end{center}
\end{table}
In table \ref{tab-goodcy4-dp} we list the distribution of del Pezzos in these ``good'' models.
\begin{table}
\begin{center}
\begin{tabular}{c|c|c|c|c|c|c|c|c|c|c|c}
{\bf class}&{\bf \# models}&{\bf \# of $dP_n$}& $dP_0$ & $dP_1$ & $dP_2$ & $dP_3$ & $dP_4$ & $dP_5$ & $dP_6$ & $dP_7$ & $dP_8$\\ 
\hline
p6v5&-&-&-&-&-&-&-&-&-&-&-\\
p7v5&$18$&$66$&$17$&$39$&-&-&$2$&-&$6$&$2$&-\\
p7v6&$57$&$212$&$39$&$92$&$10$&$7$&$7$&$5$&$26$&$21$&$5$\\
p8v5&$38$&$100$&$6$&$86$&-&-&$3$&-&$2$&$3$&-\\
p8v6&$402$&$1198$&$172$&$696$&$83$&$44$&$48$&$39$&$42$&$68$&$6$\\
p8v7&$591$&$2287$&$284$&$894$&$287$&$192$&$124$&$154$&$131$&$178$&$43$\\
p9v5&-&-&-&-&-&-&-&-&-&-&-\\
p9v6&$414$&$855$&$102$&$494$&$91$&$44$&$33$&$27$&$30$&$30$&$4$\\
p9v7&$2714$&$7378$&$902$&$3383$&$1122$&$931$&$375$&$384$&$198$&$324$&$59$\\
p9v8&$3152$&$12334$&$1377$&$4161$&$2343$&$1605$&$768$&$881$&$533$&$507$&$159$\\
\hline\hline
{\bf total}&$7386$&$24430$&$2899$&$9845$&$3936$&$2823$&$1360$&$1490$&$968$&$1133$&$276$\\
\hline
\end{tabular}\caption{Distribution of del Pezzos in ``good'' F-theory geometries.}\label{tab-goodcy4-dp}
\end{center}
\end{table}
Even if we have a reflexive fourfold polytope with nef partitions it is not implied that the nef partitions are compatible with the elliptic fibration over $B$. The extended weight systems will always lead to elliptic fibrations, but not necessarily over the base manifold we want. In many cases, there may even be more than one nef partition that is compatible with the elliptic fibration over $B$. However, these nef-partitions always lead to the same Tate model. Taking this into account we are left with $3978$ Calabi-Yau fourfolds. Our results can be found in table \ref{tab-fourfold-ellcomp}. 
\begin{table}
\begin{center}
\begin{tabular}{c|c|c}
{\bf class (base)}&{\bf \# of models w/ ell. comp. nef}&{\bf \# of ell. comp. nef }\\
\hline
p6v5&-&-\\
p7v5&$4$&$6$\\
p7v6&$46$&$83$\\
p8v5&$3$&$5$\\
p8v6&$110$&$215$\\
p8v7&$445$&$1157$\\
p9v5&-&-\\
p9v6&$69$&$116$\\
p9v7&$1014$&$2538$\\
p9v8&$2287$&$7677$\\
\hline\hline
{\bf total}&$3978$&$11797$
\end{tabular}\caption{CY fourfolds with Tate models.}\label{tab-fourfold-ellcomp}
\end{center}
\end{table}
With a nef partition in hand we can go on to construct GUT models for a particular gauge group, as described in section \ref{sec-fourfold}. For the $3978$ fourfold geometries in table \ref{tab-fourfold-ellcomp} which have a nef partition which is compatible with the elliptic fibration, we have constructed $SU(5)$ and $SO(10)$ GUT models on every del Pezzo divisor. In order to make this calculation we have to identify the coordinates of the torus fiber and the GUT divisor in the toric data of the Calabi-Yau fourfold. This can be done by matching the columns of the weight matrix of $B$ with the columns of the weight matrix of $X_4$. Note that this identification may not always be unique due to symmetries of the weight matrix. Of course, the different choices do not lead to different GUT models. One prominent example of a weight matrix with such a symmetry is the $dP_5$-model discussed in \cite{Grimm:2009yu}.\\
Carrying out this procedure we get a total number of $45\,304$ global F-theory GUTs. After removing redundancies coming from symmetries in the weight matrix, we are still left with $30\,922$ models. Note however that not all of these models will be usable, since the removal of points in the M-lattice in order to implement the GUT group may destroy the reflexivity of the polytope. In very few examples it might also happen that there is no longer a nef partition. We collect this information in table \ref{tab-gutrefl}. We make two observations: first, in about one third of the models, imposing the GUT group destroys reflexivity, and second, $U(1)$-restriction, does not put any further constraints on the reflexivity of the polytopes.
\begin{table}
\begin{center}
\begin{tabular}{c|c|c|c||c|c|c}
\hline
&\multicolumn{3}{|c||}{with redundancies}&\multicolumn{3}{|c}{without redundancies}\\
\hline
{\bf type}&{\bf refl.}&{\bf non-refl.}&{\bf no nef}&{\bf refl.}&{\bf non-refl.}&{\bf no nef}\\
\hline
$SU(5)$&$17099$&$5553$&-&$11275$&$4186$&-\\
$SO(10)$&$16625$&$6020$&$7$&$10832$&$4622$&$7$\\
$SU(5)+U(1)$-restr.&$17099$&$5553$&-&$11275$&$4186$&-\\
$SO(10)+U(1)$-restr.&$16625$&$6020$&$7$&$10832$&$4622$&$7$\\
\hline\hline
\end{tabular}\caption{Reflexivity of polytopes after implementing the GUT group.}\label{tab-gutrefl}
\end{center}
\end{table}

In the final step of our data analysis we search for new examples of F-theory GUTs which might be interesting for string phenomenology. Therefore we would like to isolate models where the GUT divisor $S$ has matter curves with a small number of moduli and not too many Yukawa points. Even though the geometries we have started with have GUT divisors with very diverse topological data, the cuts we have imposed put severe restrictions on the geometry and as a consequence also on the topological numbers of the divisors. In table \ref{tab-topdist-phys} in the appendix we list the matter genera and Yukawa points for $SU(5)$ and $SO(10)$ del Pezzos with a physical decoupling limit, and their occurrence in global models where the fourfold polytopes are reflexive after imposing the GUT group with or without $U(1)$-restriction. Similar results can be obtained for del Pezzos with a mathematical decoupling limit. 
\subsection{Examples}
We will now discuss some examples in more detail. We focus mostly on $dP_7$ and $dP_8$ since they are quite rare and $dP_8$s have not been discussed previously in the context of global models. We will also make some comments on the calculation of Euler numbers using the following formula proposed in \cite{Blumenhagen:2009yv}: given a resolved Calabi-Yau fourfold with GUT group $G$, denoted by $\bar{X}_G$, the Euler number is given by:
\begin{equation}
\label{mucformula}
\chi_{\bar{X}_G}=\chi_{\bar{X}_4}-\chi_{E_8}+\chi_{H},
\end{equation}
where $\chi_{\bar{X}_4}$ is the Euler characteristic of the resolved $X_4$ and $\chi_{H}$ denotes a correction related to $H$, which is the commutant subgroup of $G$ in $E_8$. The Euler number for a smooth elliptically fibered Calabi-Yau fourfold is:
\begin{equation}
\chi_{\bar{X}_4}=360\int_Bc_1^3(B)+12\int_Bc_1(B)c_2(B)
\end{equation}
Defining $\eta=6c_1(S)+c_1(N_S)$, the correction for $H=SU(n)$ ($n\leq 5$) is given by:
\begin{equation}
\chi_{SU(n)}=\int_S c_1^2(S)(n^3-n)+3n\eta(\eta-nc_1(S))
\end{equation}
Originally, the formula (\ref{mucformula}) was motivated from heterotic/F-theory duality and the spectral cover construction. In \cite{Grimm:2010ez} (\ref{mucformula}) has been shown to be consistent with mirror symmetry, under which $G$ and $H$ are exchanged. Note that (\ref{mucformula}) is only valid if there are no further non-abelian gauge enhancements away from the GUT brane $S$.
Furthermore, equation (\ref{mucformula}) is not valid for $U(1)$-restricted models.
In the following examples we will see that such extra enhancements can occur and lead to discrepancies in the Euler numbers of the Calabi-Yau fourfold computed by (\ref{mucformula}) and those Euler numbers obtained by PALP, which uses a formula of Batyrev and Borisov \cite{Batyrev:1995ca}. 
\subsubsection{Three $dP_8$s}
Models where the GUT divisor is a $dP_8$ are interesting for phenomenology since the genera of the matter curves and the number of Yukawa points is typically low. Unfortunately $dP_8$s are quite rare in the geometries we have constructed, and it turns out that those appearing in suitable Calabi-Yau fourfolds do not satisfy all the properties we would like to have. We will now discuss three examples. The base geometry of the first example is encoded in the following weight matrix:
\begin{equation}
\label{ex1baseweight}
\begin{tabular}{c|cccccccc|c|c}
&$y_1$&$y_2$&$y_3$&$y_4$&$y_5$&$y_6$&$y_7$&$y_8$&$\sum$&deg\\
\hline
$w_1$&$3$&$2$&$1$&$1$&$0$&$1$&$0$&$0$&$8$&$6$\\
$w_2$&$3$&$1$&$1$&$1$&$0$&$0$&$0$&$1$&$7$&$6$\\
$w_3$&$3$&$0$&$1$&$1$&$1$&$0$&$0$&$0$&$6$&$6$\\
$w_4$&$1$&$0$&$0$&$0$&$0$&$0$&$1$&$0$&$2$&$2$
\end{tabular}
\end{equation}
The second but last column indicates the sum of the weights, the last column shows the degrees of the hypersurface equation describing the base manifold $B$. In our database \cite{data} this model is labeled by {\tt (cy4)p9v6n058d6-6-6-2t1}. Let us first discuss the properties of $B$. $B$ is an almost Fano manifold and it is a Cartier divisor that is base point free. Furthermore, we only obtain three induced K\"ahler classes from the ambient space, since $D_7$ does not intersect the hypersurface, cf.~\ref{sec-induced-divisor}. There is only one del Pezzo divisor, defined by $y_6=0$, which will be our GUT divisor $S$. The topological data indicates that it is a $dP_8$. The volumes in terms of K\"ahler parameters $r_i>0$ are:
\begin{eqnarray}
\mathrm{Vol}(B)&=&6 r_1 r_2^2+2 r_2^3+36 r_1 r_2 r_3+18 r_2^2 r_3+54 r_1 r_3^2+54 r_2 r_3^2+27 r_3^3+36 r_1 r_2 r_4+18 r_2^2 r_4\nonumber\\
&&+108 r_1 r_3 r_4+108 r_2 r_3 r_4+162 r_3^2 r_4+54 r_1 r_4^2+54 r_2 r_4^2+162 r_3 r_4^2+54 r_4^3\nonumber\\
\mathrm{Vol}(S)&=&9r_3^2
\end{eqnarray}
It is easy to check that there is a mathematical as well as a physical decoupling limit. Under the mathematical decoupling limit $r_3\rightarrow 0$, $S$ is the only divisor that shrinks to zero size. If we choose $r_1\rightarrow \infty$ as a physical decoupling limit also the divisors $y_2=0$ and $y_8=0$ remain of finite size. 
However, studying this base geometry in more detail we see that it is a K3 fibration over $\mathbb P^1$. The K3 fiber degenerates at the point, $y_6=0$, of the $\mathbb P^1$ to a $dP_8$. Hence, it is a rigid divisor. Constructing a torus fibration over $B$, we observe that the coefficients $a_i$ of the fibration only depend on the coordinates of the $\mathbb P^1$. Thus, the elliptic curve remains constant over the fiber, therefore, also in the case of a degeneration. From the discriminant we find that the torus degenerates over twelve points of the $\mathbb P^1$. Hence, we obtain twelve disconnected branes along the fibers at these points and not a single connected one, as one would expect in the case of a generic fibration.

We can now na\"ively proceed and calculate the genera of the matter curves and the Yukawa numbers for a $SU(5)$ GUT on $S$. We obtain the following:
\begin{equation}
g_{SU(6)}=11\quad g_{SO(10)}=1\qquad n_{E_6}=0\quad n_{SO(12)}=0
\end{equation}
Due to the absence of Yukawa couplings this $dP_8$ is not a good candidate for a viable $SU(5)$ GUT model. However, it still can be used for an $SO(10)$ GUT where the data is as follows:
\begin{equation}
g_{SO(12)}=2\quad g_{E_6}=1\qquad n_{E_7}=2\quad n_{SO(14)}=12
\end{equation}
The weight matrix (\ref{ex1baseweight}) can be extended to a weight matrix describing a complete intersection Calabi-Yau fourfold $X_4$. The corresponding six-dimensional lattice polytope is reflexive, and there is one nef partition which respects the elliptic fibration over $B$. Using PALP we can compute the Euler number $\chi$ and the non-trivial Hodge numbers for $X_4$ and for the geometries one obtains after imposing the $SO(10)$ gauge groups. The results are collected in the following table:
\begin{equation}
\begin{tabular}{c|ccc|c}
{\bf type}&$h^{1,1}$&$h^{2,1}$&$h^{3,1}$&$\chi$\\
\hline
Tate&$12$&$26$&$54$&$288$\\
$SO(10)$&$17$&$29$&$49$&$270$\\
\end{tabular}
\end{equation}
As noticed above, already the generic fibration is rather restricted. Thus, we do not obtain 4 for $h^{1,1}$ in the unconstrained case but 12 instead. This indicates that also the $SO(10)$ results should considered with care.

For the $SO(10)$ model we can compare the Euler number to the result obtained from (\ref{mucformula}), which yields $168$. The mismatch implies that some conditions for the validity of this formula are violated. Indeed, looking at the $SU(5)/SO(10)$ Weierstrass model, we find that after imposing the GUT group on the divisor $y_6=0$, we also obtain a non-abelian enhancement on the divisor $y_8=0$. Comparing with the Tate classification, we get an $I_3^s$-enhancement for $SU(5)$ on $y_6=0$ and an $SU(3)$-enhancement for $SO(10)$. Furthermore, note that removing all the monomials in the Weierstrass equation, that do not comply with $SU(5)/SO(10)$, the $(a_0,a_1,a_2,a_3,a_6)$ schematically (i.e. after setting all complex structure parameters to $1$) vanish as follows on $S$: $(1+w^2,w^2+w^4,w^2+w^4+w^6,w^4+w^6+w^8,w^6+w^8+\ldots)$ for $SU(5)$, and $(w^2,w^2+w^4,w^2+w^4+w^6,w^4+w^6+w^8,w^6+w^8+\ldots)$ for $SO(10)$. Thus, the singularity enhancements are actually higher than that of $SU(5)$ or $SO(10)$. As we observed already above, the reason for all the problems roots in the  very ungeneric form of the coefficients in the Weierstrass model. This comes from the fact that the anti-canonical class does not depend on all toric classes. We see that constructing a Tate model over a promising base manifold may not lead to the wanted brane setup.\\\\
As indicated in table \ref{tab-topdist-phys} the $dP_8$ with the matter genera and Yukawa numbers above is the only one with a physical decoupling limit. The $dP_8$s we have found in the global models we have constructed only have very few combinations of topological numbers. In order to also give an example where an $SU(5)$ GUT is possible, we consider the following base geometry:
\begin{equation}
\label{ex2baseweight}
\begin{tabular}{c|cccccccc|c|c}
&$y_1$&$y_2$&$y_3$&$y_4$&$y_5$&$y_6$&$y_7$&$y_8$&$\sum$&deg\\
\hline
$w_1$&$1$&$1$&$0$&$0$&$0$&$0$&$0$&$0$&$2$&$1$\\
$w_2$&$1$&$0$&$1$&$0$&$1$&$0$&$1$&$0$&$4$&$3$\\
$w_3$&$1$&$0$&$1$&$0$&$0$&$1$&$0$&$1$&$4$&$3$\\
$w_4$&$0$&$0$&$1$&$1$&$1$&$0$&$0$&$0$&$3$&$2$
\end{tabular}
\end{equation}
The file name in the database is {\tt (cy4)p9v8n224d1-3-3-2t1}. As in the previous examples the base $B$ is almost Fano. The hypersurface divisor is Cartier and base point free. There are two del Pezzo divisors, one $dP_8$ and one $dP_5$. We focus on the $dP_8$ here, which is given by $y_1=0$. The volumes of $B$ and $S$ are:
\begin{eqnarray}
\mathrm{Vol}(B)&=&2 r_1^3+15 r_1^2 r_2+6 r_1 r_2^2+18 r_1^2 r_3+30 r_1 r_2 r_3+6 r_2^2 r_3+18 r_1 r_3^2+15 r_2 r_3^2+6 r_3^3\nonumber\\
&&+18 r_1^2 r_4+30 r_1 r_2 r_4+6 r_2^2 r_4+48 r_1 r_3 r_4+30 r_2 r_3 r_4+24 r_3^2 r_4+24 r_1 r_4^2\nonumber\\
&&+15 r_2 r_4^2+24 r_3 r_4^2+8 r_4^3\nonumber\\
\mathrm{Vol}(S)&=&(r_1+r_3+r_4)(5(r_1+r_3+r_4)+4\,r_2)
\end{eqnarray}
Clearly, there is no decoupling limit. This can also be seen from the fact that $S$ is not a rigid divisor. $B$ is a $\mathbb P^1$ fibration over a toric $dP_1$ and $S$ the reduction of this fibration over a non-rigid curve in this $dP_1$.

Computing the matter genera and the Yukawa numbers one finds for $SU(5)$:
\begin{equation}
g_{SU(6)}=74\quad g_{SO(10)}=2\qquad n_{E_6}=8\quad n_{SO(12)}=11
\end{equation}
and for $SO(10)$:
\begin{equation}
g_{SO(12)}=9\quad g_{E_6}=5\qquad n_{E_7}=16\quad n_{SO(14)}=52
\end{equation}
The fourfold $X_4$ is described by a reflexive polyhedron with $17$ nef partitions, four of which describe an elliptic fibration over $B$. The Hodge numbers are collected in the table below:
\begin{equation}
\begin{tabular}{c|ccc|c}
{\bf type}&$h^{1,1}$&$h^{2,1}$&$h^{3,1}$&$\chi$\\
\hline
Tate&$5$&$9$&$404$&$2448$\\
$SU(5)$&$13$&$9$&$84$&$360$\\
$SO(10)$&$17$&$11$&$43$&$342$\\
$SU(5)_{U(1)}$&$14$&$9$&$44$&$342$\\
$SO(10)_{U(1)}$&$18$&$11$&$39$&$324$
\end{tabular}
\end{equation}
Again, the Hodge numbers for $SU(5)$/$SO(10)$, without $U(1)$-restriction, do note fit the numbers calculated with formula (\ref{mucformula}).
Examining the Tate equation after imposing the GUT group, we find an additional gauge enhancement at the divisor $y_4=0$. For $SU(5)$ the extra enhancement is also $SU(5)$, for $SO(10)$, the $y_4=0$ also carries an $SO(10)$ enhancement. Note that the second del Pezzo divisor in $B$, $y_2=0$, which is a $dP_5$ has a mathematical and a physical decoupling limit. It is a rigid divisor and the Euler numbers after imposing the GUT groups on it match the Euler numbers computed with~\eqref{mucformula}. The form of the Tate equation implies that in that case no other divisor gets a non-abelian enhancement.\\\\
Finally, we consider an example of a $dP_8$ with a mathematical decoupling limit. The base geometry is given by the following weight matrix:
\begin{equation}
\label{ex2abaseweight}
\begin{tabular}{c|ccccccc|c|c}
&$y_1$&$y_2$&$y_3$&$y_4$&$y_5$&$y_6$&$y_7$&$\sum$&deg\\
\hline
$w_1$&$1$&$1$&$0$&$0$&$0$&$0$&$0$&$2$&$2$\\
$w_2$&$1$&$0$&$1$&$1$&$1$&$0$&$0$&$4$&$3$\\
$w_3$&$2$&$0$&$1$&$1$&$0$&$1$&$1$&$6$&$5$
\end{tabular}
\end{equation}
In the database this model is labeled by {\tt (cy4)p8v7n073d2-3-5t1}. 
There are two del Pezzo divisors: $y_2=0$ is a $dP_0$ and $y_5=0$, which we will name $S$, is $dP_8$. The existence of a mathematical decoupling limits can be deduced from the volumes of the base $B$ and $S$:
\begin{eqnarray}
\mathrm{Vol}(B)&=&2 r_1^3+15 r_1^2 r_2+24 r_1 r_2^2+11 r_2^3+15 r_1^2 r_3+60r_1 r_2 r_3+48 r_2^2 r_3+30 r_1 r_3^2+60 r_2 r_3^2+20 r_3^3\nonumber\\
\mathrm{Vol}(S)&=&4 r_1 r_2+5 r_2^2+8 r_2 r_3
\end{eqnarray}
The mathematical decoupling limit can be implemented by setting $r_2\rightarrow 0$. In that case none of the other divisors will shrink to zero size. 
The topological data of the matter curves and the Yukawa couplings for $SU(5)$-models is:
\begin{equation}
g_{SU(6)}=38\quad g_{SO(10)}=0\qquad n_{E_6}=2\quad n_{SO(12)}=4
\end{equation}
and for $SO(10)$:
\begin{equation}
g_{SO(12)}=5\quad g_{E_6}=2\qquad n_{E_7}=8\quad n_{SO(14)}=32
\end{equation}
Two nef partitions are compatible with the elliptic fibration. The Hodge numbers and the Euler number are collected in the following table:
\begin{equation}
\begin{tabular}{c|ccc|c}
{\bf type}&$h^{1,1}$&$h^{2,1}$&$h^{3,1}$&$\chi$\\
\hline
Tate&$4$&$26$&$182$&$1008$\\
$SU(5)$&$8$&$26$&$83$&$438$\\
$SO(10)$&$9$&$26$&$81$&$432$\\
$SU(5)_{U(1)}$&$9$&$26$&$71$&$372$\\
$SO(10)_{U(1)}$&$10$&$26$&$69$&$366$
\end{tabular}
\end{equation}
Even though there are no further non-abelian enhancements on the torically induced divisors of $B$, the Euler numbers do not match those obtained from (\ref{mucformula}). The mismatch might still be due to an extra non-abelian enhancement on a divisor which is not toric. 
Another possible explanation could be that we have a non-abelian enhancement over a curve. Resolving the singularities on these curves leads to a further K\"ahler parameter. However, we do not observe the corresponding K\"ahler modulus in the above table.

\subsubsection{Three $dP_7$s}
As a second class of examples we discuss a model which has two different $dP_7$ divisors. The base is specified by the following weight matrix and hypersurface degrees:
\begin{equation}
\label{ex3baseweight}
\begin{tabular}{c|cccccccc|c|c}
&$y_1$&$y_2$&$y_3$&$y_4$&$y_5$&$y_6$&$y_7$&$y_8$&$\sum$&deg\\
\hline
$w_1$&$1$&$1$&$0$&$0$&$0$&$0$&$0$&$0$&$2$&$2$\\
$w_2$&$1$&$0$&$1$&$1$&$0$&$1$&$0$&$1$&$5$&$4$\\
$w_3$&$1$&$0$&$0$&$0$&$1$&$1$&$0$&$0$&$3$&$2$\\
$w_4$&$0$&$0$&$1$&$1$&$0$&$0$&$1$&$0$&$3$&$2$
\end{tabular}
\end{equation}
The identifier for this model is {\tt (cy4)p9v8n152d2-4-2-2t2}. 
The two $dP_7$s are given by $y_5=0$ and $y_7=0$, and we call the associated GUT branes $S_5$ and $S_7$. 
Let us first discuss the decoupling limits.
\begin{eqnarray}
\mathrm{Vol}(B)&=&6 r_1^2 r_2+6 r_1 r_2^2+2 r_2^3+6 r_1^2 r_3+24 r_1 r_2 r_3+12 r_2^2 r_3+6 r_1 r_3^2+6 r_2 r_3^2+6 r_1^2 r_4+24 r_1 r_2 r_4\nonumber\\
&&+12 r_2^2 r_4+24 r_1 r_3 r_4+24 r_2 r_3 r_4+6 r_3^2r_4+12 r_1 r_4^2+12 r_2 r_4^2+12 r_3 r_4^2+4 r_4^3\nonumber\\
\mathrm{Vol}(S_5)&=&2 r_1^2+4 r_1 r_3+4 r_1 r_4\nonumber \\
\mathrm{Vol}(S_7)&=&4 r_1 r_3+4 r_2 r_3+2 r_3^2+4 r_3 r_4
\end{eqnarray}
As can be easily verified, $S_5$ has a mathematical as well as a physical decoupling limit, whereas $S_7$ only has a mathematical decoupling limit. The K\"ahler parameters can always be chosen in such a way that the respective GUT divisor is the only one whose volume goes to zero/remains finite in the mathematical/physical decoupling limit. The matter genera and Yukawa numbers for $S_5$ are the following:
\begin{equation}
g_{SU(6)}=21\quad g_{SO(10)}=1\qquad n_{E_6}=0\quad n_{SO(12)}=0
\end{equation}
for $SU(5)$ and
\begin{equation}
g_{SO(12)}=3\quad g_{E_6}=1\qquad n_{E_7}=4\quad n_{SO(14)}=24
\end{equation}
for $SO(10)$. As in the first $dP_8$-example, $S_5$ is not suitable for $SU(5)$ GUTs due to the absence of Yukawa points. For the divisor $S_7$ the topological data for $SU(5)$ and $SO(10)$ GUTs are as follows:
\begin{equation}
g_{SU(6)}=48\quad g_{SO(10)}=0\qquad n_{E_6}=2\quad n_{SO(12)}=4
\end{equation}
for $SU(5)$ and
\begin{equation}
g_{SO(12)}=6\quad g_{E_6}=2\qquad n_{E_7}=10\quad n_{SO(14)}=44
\end{equation}
for $SO(10)$.
The associated Calabi-Yau fourfold has $25$ nef partitions, three of which describe an elliptic fibration over $B$. Imposing the GUT groups on $S_5$ (first block) and $S_7$ (second block) we compute the following Hodge numbers:
\begin{equation}
\begin{tabular}{c|ccc|c}
{\bf type}&$h^{1,1}$&$h^{2,1}$&$h^{3,1}$&$\chi$\\
\hline
Tate&$5$&$11$&$1066$&$1008$\\
$SU(5)$&$9$&$10$&$121$&$768$\\
$SO(10)$&$10$&$10$&$120$&$768$\\
$SU(5)_{U(1)}$&$10$&$10$&$78$&$516$\\
$SO(10)_{U(1)}$&$11$&$10$&$77$&$516$\\
\hline
$SU(5)$&$9$&$11$&$67$&$438$\\
$SO(10)$&$10$&$11$&$65$&$432$\\
$SU(5)_{U(1)}$&$10$&$11$&$55$&$372$\\
$SO(10)_{U(1)}$&$11$&$11$&$53$&$366$\\
\end{tabular}
\end{equation}
For the $SU(5)$ and $SU(10)$ model on $S_5$ the Euler numbers agree with the formula (\ref{mucformula}) of \cite{Blumenhagen:2009yv}, and there are also no further non-abelian enhancements in the Tate models. For $S_7$ there is a mismatch of Euler numbers, even though we do not find any further non-abelian enhancements on the toric divisors of $B$ in the Tate model. However, there may be some singularities over non-toric divisors.
\\\\
Now we would like to discuss a $dP_7$-model with a physical decoupling limit. For this purpose we look at a base geometry which is specified by the following data:
\begin{equation}
\label{ex4baseweight}
\begin{tabular}{c|cccccccc|c|c}
&$y_1$&$y_2$&$y_3$&$y_4$&$y_5$&$y_6$&$y_7$&$y_8$&$\sum$&deg\\
\hline
$w_1$&$1$&$1$&$0$&$0$&$0$&$0$&$0$&$0$&$2$&$2$\\
$w_2$&$1$&$0$&$1$&$1$&$0$&$0$&$0$&$0$&$3$&$2$\\
$w_3$&$3$&$0$&$0$&$1$&$0$&$1$&$1$&$1$&$7$&$6$\\
$w_4$&$2$&$0$&$0$&$0$&$1$&$1$&$1$&$0$&$5$&$4$
\end{tabular}
\end{equation}
In the database the label of this model is {\tt (cy4)p9v8n341d2-2-6-4t1}. 
This model also has two $dP_7$s given by $y_3=0$ and $y_4=0$. The former has the same matter genera and Yukawa points as $S_5$ above, so we will focus on the latter which we will call $S$. The divisor $S$ is not rigid. To see this we have to examine $B$ in more detail. We find that $B$ is a $dP_7$ fibration over $\mathbb{P}^1$. Furthermore, the typical fiber of this fibration is equivalent to $S$. We note further that the divisor $D_2$ of the ambient space does not intersect the hypersurface, cf.~section~\ref{sec-induced-divisor}. The existence of a physical decoupling limit is inferred from the volumes of $B$ and $S$:
\begin{eqnarray}
\mathrm{Vol}(B)&=&6 r_1 r_2^2+2 r_2^3+24 r_1 r_2 r_3+18 r_2^2 r_3+24 r_1 r_3^2+48 r_2 r_3^2+24 r_3^3+24 r_1 r_2 r_4+18 r_2^2 r_4\nonumber\\
&&+48 r_1 r_3 r_4+96 r_2 r_3 r_4+120 r_3^2 r_4+24 r_1 r_4^2+48r_2 r_4^2+120 r_3 r_4^2+40 r_4^3\nonumber\\
\mathrm{Vol}(S)&=&2 (r_2+2\, r_3+2\, r_4)^2
\end{eqnarray}
The physical decoupling limit is achieved when we take $r_1\rightarrow\infty$ which is the volume of the $\mathbb{P}^1$, the base space of the fibration. In this limit also the other $dP_7$ $y_3=0$, which is also a fiber, remains of finite size. 
The matter and Yukawa data for $SU(5)$ and $SO(10)$ GUTs are:
\begin{equation}
g_{SU(6)}=57\quad g_{SO(10)}=1\qquad n_{E_6}=4\quad n_{SO(12)}=6
\end{equation}
for $SU(5)$ and
\begin{equation}
g_{SO(12)}=7\quad g_{E_6}=3\qquad n_{E_7}=12\quad n_{SO(14)}=48
\end{equation}
for $SO(10)$.
Extending the weight matrix of the base manifold we get an elliptically fibered Calabi-Yau fourfold which has $7$ nef partitions. Three of these are elliptic fibrations over $B$ as given by (\ref{ex4baseweight}). Computing the Hodge data, we get the following results:
\begin{equation}
\begin{tabular}{c|ccc|c}
{\bf type}&$h^{1,1}$&$h^{2,1}$&$h^{3,1}$&$\chi$\\
\hline
Tate&$4$&$22$&$178$&$1008$\\
$SU(5)$&$12$&$22$&$53$&$306$\\
$SO(10)$&$16$&$24$&$50$&$300$\\
$SU(5)_{U(1)}$&$13$&$22$&$51$&$300$\\
$SO(10)_{U(1)}$&$17$&$24$&$48$&$294$\\
\end{tabular}
\end{equation}
Again, the Euler number from the Hodge data disagree with the one calculated from formula~\eqref{mucformula}. Looking at the Tate model for the F-theory GUT we find an additional $SU(5)$ or $SO(10)$-enhancement on the divisor $y_5=0$.
\subsubsection{The toric three-/fourfold of~\cite{Marsano:2009ym}}
The last example that we consider is the model {\tt (cy4)p8v7n080d1-1-3t1}, which is equivalent to the compactification geometry discussed in~\cite{Marsano:2009ym}, cf. also~\cite{diss}. The base geometry is given by the following weight matrix and hypersurface:
\begin{equation}
\label{ex5baseweight}
\begin{array}{c|ccccccc|c|c}
&y_1 & y_2 & y_3 & y_4 & y_5 & y_6 & y_7 & \sum & \textmd{deg}\\
\hline
 w_1 & 1 & 1 & 0 & 0 & 0 & 0 & 0 & 2 & 1 \\
 w_2 & 0 & 1 & 1 & 1 & 0 & 0 & 0 & 3 & 1 \\
 w_3 & 0 & 2 & 1 & 0 & 1 & 1 & 1 & 6 & 3 
\end{array}\,.
\end{equation}
This is an example of a base manifold which does not satisfy the almost Fano condition. 
The relevant $dP_2$ on which we will place the GUT model is $D_4$. Together with the $dP_1$ on $D_1$, these are the only two shrinkable del Pezzo surfaces as one can see from the volumes of $B$, $S=D_4$, and $D_1$,
\begin{eqnarray}
\textmd{Vol}&=& 3 r_1^2 r_2+3 r_1 r_2^2+r_2^3+3 r_1^2 r_3+18 r_1 r_2 r_3+9 r_2^2 r_3+12 r_1 r_3^2+18 r_2 r_3^2+10 r_3^3\nonumber\\
\textmd{S}&=& (r_1+2 r_3)^2\\
D_1&=& r_2(2\, r_1+r_2)\nonumber\,.
\end{eqnarray}
Besides these two rigid del Pezzos there are other toric $dP_2$s on $D_5\sim D_6\sim D_7$ which do not have a decoupling limit. Before we come to the fourfold geometry, we compute the matter and Yukawa data for $SU(5)$ and $SO(10)$ GUTs on $S$:
\begin{equation}
g_{SU(6)}=134 \quad g_{SO(10)}=0 \qquad n_{E_6}= 6\quad n_{SO(12)}= 10
\end{equation}
for $SU(5)$ and
\begin{equation}
g_{SO(12)}= 15\quad g_{E_6}= 4\qquad n_{E_7}= 28 \quad n_{SO(14)}= 128
\end{equation}
for $SO(10)$.

Again, we extend  the weight matrix of the base manifold to obtain an elliptically fibered Calabi-Yau fourfold which has $4$ nef partitions. Two of these are elliptic fibrations over $B$ as given by (\ref{ex5baseweight}). Computing the Hodge data for this fourfold and the reduced ones, we obtain the following results:
\begin{equation}
\begin{array}{c|ccc|c}
\textmd{\bf type}& h^{1,1} & h^{2,1} & h^{3,1} & \chi \\
\hline
\textmd{Tate}  &  4  &  0  &  2316  & 13968   \\
 SU(5)         &  8  &  0  &  1867  & 11298  \\
 SO(10)        &  9  &  0  &  1863  & 11280 \\
 SU(5)_{U(1)}  &  9  &  0  &   796  & 4878 \\
 SO(10)_{U(1)} & 10  &  0  &   792  & 4860
\end{array}\,.
\end{equation}
These results match with the outcome of $SU(5)$/$SO(10)$ one finds from \eqref{mucformula}.


\section{Conclusions}
\label{sec-conc}
In this paper we have constructed a large class of Calabi-Yau fourfolds which are particularly useful for F-theory compactifications. There are several interesting directions for continued research. 

Having such a large class of examples it might be useful to extend the rather basic analysis and to do more detailed calculations in F-theory. One possibility would be to include calculations with fluxes. It has been argued in \cite{Blumenhagen:2009yv,Grimm:2010ez,Marsano:2010ix,Chung:2010bn} that the spectral cover construction which can be used to describe fluxes locally near the GUT brane \cite{Donagi:2009ra} is valid in certain cases also beyond the local picture. Our data contains all the input needed to calculate chiral indices and tadpole cancellation conditions for a large class of models. Also the flux quantization and anomaly cancellation conditions worked out in \cite{Marsano:2010sq,Collinucci:2010gz} could be included into the analysis. 

In \cite{Braun:2010hr} F-theory models where the GUT brane does not wrap a del Pezzo divisor have been discussed. Despite the fact that the connection to many of the local GUT models discussed in the literature is not immediate, these GUTs are interesting because they may allow for gauge group breaking by discrete Wilson lines. The analysis we have performed for del Pezzo divisors can be extended to toric divisors in the base which are not del Pezzo.

So far no examples have been discussed where it is possible to make contact between F-theory GUT models and the Calabi-Yau fourfolds which are encountered in the calculation of $N=1$ superpotentials \cite{Alim:2009bx,Grimm:2009ef,Grimm:2009sy,Jockers:2009ti,Alim:2010za,Grimm:2010gk}. One could search our database for fourfold geometries which are suited for establishing a connection between these two exciting topics.

In our calculations we have made use of an extension of the software package PALP \cite{maxnils}, which can compute triangulations of polytopes and calculates the Mori cone, the Stanley-Reisner ideal and intersection rings for hypersurfaces in toric ambient spaces. An extension of these routines to the case of complete intersection Calabi-Yaus is interesting not only for applications in F-theory GUTs. Furthermore it would be useful to extend PALP to handle also non-reflexive polytopes. In this context the program cohomCalg \cite{Blumenhagen:2010pv} may be helpful for the calculation of Hodge numbers. Finally we should also try to overcome the problems with numerical overflows that arose due to the complexity of the fourfold polytopes.

A more mathematical question concerns methods to partially classify Calabi-Yau fourfolds. A complete classification of Calabi-Yau fourfolds that are hypersurfaces or complete intersections in a toric ambient space seems to be out of reach. An empirical formula due to H. Skarke \cite{harald} estimates the number $N_d$ of reflexive polytopes in $d$ dimensions to be of order $N_d\simeq 2^{2^{d+1}-4}$.  This implies that the number of reflexive polytopes in $5$ dimension is of order $\mathcal{O}(10^{18})$. In $6$ dimensions there are even expected to be $\mathcal{O}(10^{37})$ reflexive polytopes. Since also non-reflexive polytopes may be of interest in F-theory, this number might only be the tip of the iceberg. Even a classification of elliptically fibered Calabi-Yau fourfolds may be too difficult. However, what could in principle be doable is a complete classification of the geometries we have constructed in this article. The prescription is the following:  take each of the $473\,800\,776$ reflexive polyhedra in four dimensions and put in all possible hypersurfaces whose degree is below the degree of the Calabi-Yau hypersurface in this ambient space. Then construct fourfolds which are elliptic fibrations over these base manifolds. A na\"ive estimate shows that this procedure would yield $\mathcal{O}(10^{11})$ fourfold geometries. Due to the overflow problem we can only claim that we have a full classification of this type of Calabi-Yau fourfolds if they originate from reflexive polyhedra in four dimensions which have up to seven points.


\appendix
\section{Matter Genera and Yukawa Points}
In the following table we list the matter genera and Yukawa numbers for those del Pezzos, where the F-theory GUT lives on a Calabi-Yau fourfold described by to a reflexive polytope, where at least one nef partition is compatible with the elliptic fibration. Furthermore, the base $B$ should be regular, and at least one of the del Pezzos inside the base should admit a decoupling limit. Note that for the calculation of these numbers the formulas (\ref{eq:euler-number-curves}) and (\ref{nyuk}) have been used. There it was assumed that the curves involved are irreducible. Since we could not check this explicitly for every model, some of these numbers might be incorrect.
\begin{center}
\begin{longtable}{c||c|c|c|c||c|c|c|c||c}
\hline
&\multicolumn{4}{|c||}{$SU(5)$}&\multicolumn{4}{|c||}{$SO(10)$}&\\
\hline
{\bf type}&{\bf $g_{SU(6)}$}&{\bf $g_{SO(10)}$}&{\bf $n_{E_6}$}&{\bf $n_{SO(12)}$}&{\bf $g_{SO(12)}$}&{\bf $g_{E_6}$}&{\bf $n_{E_7}$}&{\bf $n_{SO(14)}$}&{\bf \#}\\
\hline\hline
$dP_8$&$11$&$1$&$0$&$0$&$2$&$1$&$2$&$12$&$9$\\
\hline
$dP_7$&$57$&$1$&$4$&$6$&$7$&$3$&$12$&$48$&$187$\\
&$102$&$2$&$10$&$14$&$12$&$6$&$22$&$76$&$2$\\
&$75$&$1$&$6$&$9$&$9$&$4$&$16$&$60$&$5$\\
&$21$&$1$&$0$&$0$&$3$&$1$&$4$&$24$&$73$\\
&$48$&$0$&$2$&$4$&$6$&$2$&$10$&$44$&$1$\\
&$66$&$0$&$4$&$7$&$8$&$3$&$14$&$56$&$2$\\
\hline
$dP_6$&$85$&$1$&$6$&$9$&$10$&$4$&$18$&$72$&$161$\\
&$31$&$1$&$0$&$0$&$4$&$1$&$6$&$36$&$47$\\
&$58$&$0$&$2$&$4$&$7$&$2$&$12$&$56$&$32$\\
&$130$&$2$&$12$&$17$&$15$&$7$&$28$&$100$&$3$\\
&$76$&$0$&$4$&$7$&$9$&$3$&$16$&$68$&$4$\\
&$103$&$1$&$8$&$12$&$12$&$5$&$22$&$84$&$3$\\
\hline
$dP_5$&$68$&$0$&$2$&$4$&$8$&$2$&$14$&$68$&$96$\\
&$113$&$1$&$8$&$12$&$13$&$5$&$24$&$96$&$340$\\
&$104$&$0$&$6$&$10$&$12$&$4$&$22$&$92$&$7$\\
&$131$&$1$&$10$&$15$&$15$&$6$&$28$&$108$&$14$\\
&$158$&$2$&$14$&$20$&$18$&$8$&$34$&$124$&$17$\\
&$86$&$0$&$4$&$7$&$10$&$3$&$18$&$80$&$34$\\
&$41$&$1$&$0$&$0$&$5$&$1$&$8$&$48$&$47$\\
&$185$&$3$&$18$&$25$&$21$&$10$&$40$&$140$&$3$\\
&$176$&$2$&$16$&$23$&$20$&$9$&$38$&$136$&$1$\\
\hline
$dP_4$&$141$&$1$&$10$&$15$&$16$&$6$&$30$&$120$&$141$\\
&$96$&$0$&$4$&$7$&$11$&$3$&$20$&$92$&$56$\\
&$78$&$0$&$2$&$4$&$9$&$2$&$16$&$80$&$60$\\
&$186$&$2$&$16$&$23$&$21$&$9$&$40$&$148$&$16$\\
&$51$&$1$&$0$&$0$&$6$&$1$&$10$&$60$&$21$\\
&$114$&$0$&$6$&$10$&$13$&$4$&$24$&$104$&$23$\\
&$159$&$1$&$12$&$18$&$18$&$7$&$34$&$132$&$4$\\
&$132$&$0$&$8$&$13$&$15$&$5$&$28$&$116$&$10$\\
\hline
$dP_3$&$124$&$0$&$6$&$10$&$14$&$4$&$26$&$116$&$189$\\
&$169$&$1$&$12$&$18$&$19$&$7$&$36$&$144$&$267$\\
&$205$&$1$&$16$&$24$&$23$&$9$&$44$&$169$&$28$\\
&$160$&$0$&$10$&$16$&$18$&$6$&$34$&$140$&$6$\\
&$268$&$4$&$26$&$36$&$30$&$14$&$58$&$204$&$10$\\
&$214$&$2$&$28$&$26$&$24$&$10$&$46$&$172$&$18$\\
&$88$&$0$&$2$&$4$&$10$&$2$&$28$&$92$&$63$\\
&$61$&$1$&$0$&$0$&$71$&$1$&$12$&$72$&$32$\\
&$142$&$0$&$8$&$13$&$16$&$5$&$30$&$128$&$45$\\
&$106$&$0$&$4$&$7$&$12$&$3$&$22$&$104$&$35$\\
&$187$&$1$&$14$&$21$&$21$&$8$&$40$&$156$&$15$\\
&$250$&$2$&$22$&$32$&$28$&$12$&$54$&$196$&$1$\\
&$241$&$3$&$22$&$31$&$27$&$12$&$52$&$188$&$5$\\
\hline
$dP_2$&$170$&$0$&$10$&$16$&$19$&$6$&$36$&$152$&$218$\\
&$134$&$0$&$6$&$10$&$15$&$4$&$28$&$128$&$180$\\
&$197$&$1$&$14$&$21$&$22$&$8$&$42$&$168$&$427$\\
&$215$&$1$&$16$&$24$&$24$&$9$&$46$&$180$&$102$\\
&$269$&$3$&$24$&$34$&$30$&$13$&$58$&$212$&$25$\\
&$116$&$7$&$4$&$7$&$13$&$3$&$24$&$116$&$73$\\
&$242$&$2$&$20$&$29$&$27$&$11$&$52$&$196$&$105$\\
&$188$&$0$&$12$&$19$&$21$&$7$&$40$&$164$&$18$\\
&$71$&$1$&$0$&$0$&$8$&$1$&$14$&$84$&$30$\\
&$152$&$0$&$8$&$13$&$17$&$5$&$32$&$140$&$117$\\
&$260$&$2$&$22$&$32$&$29$&$12$&$56$&$208$&$22$\\
&$323$&$5$&$32$&$44$&$36$&$17$&$70$&$244$&$10$\\
&$98$&$0$&$2$&$4$&$11$&$2$&$20$&$104$&$34$\\
&$296$&$4$&$28$&$39$&$33$&$15$&$64$&$228$&$19$\\
&$206$&$0$&$14$&$22$&$23$&$8$&$44$&$176$&$1$\\
&$305$&$3$&$28$&$40$&$34$&$15$&$66$&$236$&$2$\\
\hline
$dP_1$&$225$&$1$&$16$&$24$&$25$&$9$&$48$&$192$&$1150$\\
&$252$&$0$&$18$&$28$&$28$&$10$&$54$&$212$&$11$\\
&$144$&$0$&$6$&$10$&$16$&$4$&$30$&$140$&$482$\\
&$81$&$1$&$0$&$0$&$9$&$1$&$16$&$96$&$214$\\
&$198$&$0$&$12$&$19$&$22$&$7$&$42$&$176$&$139$\\
&$270$&$2$&$22$&$32$&$30$&$12$&$58$&$220$&$239$\\
&$180$&$0$&$10$&$16$&$20$&$6$&$38$&$164$&$603$\\
&$162$&$0$&$8$&$13$&$18$&$5$&$34$&$152$&$476$\\
&$315$&$3$&$28$&$40$&$35$&$15$&$68$&$248$&$54$\\
&$378$&$6$&$38$&$52$&$42$&$20$&$82$&$284$&$20$\\
&$108$&$0$&$2$&$4$&$12$&$2$&$22$&$116$&$278$\\
&$441$&$9$&$48$&$64$&$49$&$25$&$96$&$320$&$9$\\
&$234$&$0$&$16$&$25$&$26$&$9$&$50$&$200$&$7$\\
&$243$&$1$&$18$&$27$&$27$&$10$&$52$&$204$&$51$\\
&$216$&$0$&$14$&$22$&$24$&$8$&$46$&$188$&$28$\\
&$297$&$3$&$26$&$37$&$33$&$14$&$64$&$236$&$54$\\
&$324$&$4$&$30$&$42$&$36$&$16$&$70$&$252$&$27$\\
&$126$&$0$&$4$&$7$&$14$&$3$&$26$&$128$&$175$\\
&$351$&$5$&$34$&$47$&$39$&$18$&$76$&$268$&$15$\\
&$270$&$0$&$20$&$31$&$30$&$11$&$58$&$224$&$1$\\
\hline
$dP_0$&$253$&$1$&$18$&$27$&$28$&$10$&$54$&$216$&$338$\\
&$496$&$10$&$54$&$72$&$55$&$28$&$108$&$360$&$12$\\
&$91$&$1$&$0$&$0$&$10$&$1$&$18$&$108$&$150$\\
&$190$&$0$&$10$&$16$&$21$&$6$&$40$&$176$&$763$\\
&$325$&$3$&$28$&$40$&$36$&$15$&$70$&$260$&$126$\\
&$136$&$0$&$4$&$7$&$15$&$3$&$28$&$140$&$380$\\
&$406$&$6$&$40$&$55$&$45$&$21$&$88$&$308$&$33$\\
\hline\hline
\caption{Topological numbers of del Pezzos with physical decoupling limit.}\label{tab-topdist-phys}
\end{longtable}
\end{center}

\phantomsection
\addcontentsline{toc}{section}{References}
\bibliographystyle{fullsort}
\bibliography{proj_gftm_papers.bib}

\end{document}